\documentclass[preprint,3p,12pt]{elsarticle}
\usepackage{verbatim}
\usepackage{graphicx}
\usepackage{float} 
\usepackage{subcaption}%
\def\BibTeX{{\rm B\kern-.05em{\sc i\kern-.025em b}\kern-.08em
		T\kern-.1667em\lower.7ex\hbox{E}\kern-.125emX}}

\usepackage{balance}

\usepackage{diagbox}

\usepackage{bm}
\usepackage{amsmath}
\usepackage{amssymb}
\usepackage{epsfig}
\usepackage{multirow}
\usepackage{xcolor}

\def\bS{{\bf{S}}}
\def\bX{{\bf{X}}}

\def\bu{{\bf{u}}}
\def\bv{{\bf{v}}}

\def\bT{{\bf{\Theta}}}

\def\be{\begin{equation}}
\def\en{\end{equation}}

\DeclareMathOperator{\E}{\mathbb{E}}

\def\RR{\rm \hbox{I\kern-.2em\hbox{R}}}
\def\NN{\rm \hbox{I\kern-.2em\hbox{N}}}
\def\ZZ{\rm {{Z}\kern-.28em{Z}}}
\def\CC{\rm \hbox{C\kern -.5em {\raise .32ex \hbox{$\scriptscriptstyle
				|$}}\kern
		-.22em{\raise .6ex \hbox{$\scriptscriptstyle |$}}\kern .4em}}
\def\L{\pounds}
\def\<{\langle}
\def\>{\rangle}

\long\def\symbolfootnote[#1]#2{\begingroup%
	\def\thefootnote{\fnsymbol{footnote}}\footnote[#1]{#2}\endgroup}

\begin{document}
	
\begin{frontmatter}	
\title{Improving probabilistic wind speed forecasting using M-Rice distribution and spatial data integration}

\author[]{Roberta Baggio\corref{cor1}}
\ead{baggio_r@univ-corse.fr}
\author[]{Jean-Fran\c{c}ois Muzy}
\ead{muzy_j@univ-corse.fr}
\address{\footnotesize \em Laboratoire ``Sciences Pour l'Environnement''  \\	\footnotesize \em UMR 6134 CNRS - University of Corsica \\
\footnotesize \em Campus Grimaldi, 20250 Corte (France)}

\cortext[cor1]{Corresponding author}

			
\begin{abstract}
We consider the problem of short-term forecasting of surface wind speed probability distribution. Our approach simply consists in predicting the parameters 	of a given probability density function by training a neural network model whose loss function is chosen as the log-likelihood provided by this distribution. We compare different possibilities among a set of distributions that have been previously considered in the context of modeling wind fluctuations. Our results rely on two different hourly wind speed datasets: the first one has been recorded by M\'et\'eo-France in Corsica (South France), a very mountainous Mediterranean island while the other one relies on KNMI database that provides records of various stations over the Netherlands, a very flat country in Northwestern Europe. 
	A first part of our work  globally unveils the superiority of the so-called "Multifractal Rice" (M-Rice) distribution over alternative parametric models, showcasing its potential as a reliable tool for wind speed forecasting.  This family of distributions has been proposed  in the context of modeling wind speed fluctuations as a random cascade model along the same picture as fully developed turbulence. For all stations in both regions, it consistently provides better results regardless of the considered probabilistic scoring rule or forecasting horizon. Our second findings demonstrate significant enhancements in forecasting accuracy when one incorporates wind speed data from proximate weather stations, in full  agreement with the results obtained formerly for point-wise wind speed prediction. Moreover, we reveal that the incorporation of ERA5 reanalysis of 10 m wind data from neighboring grid points contributes to a substantial improvement at time horizon $h=6$ hours  while at a shorter time horizon ($h=1$ h) it does not lead to any noticeable improvement.  It turns out that accounting for pertinent features and explanatory factors, notably those related the spatial distribution and wind speed and direction, emerges as a more critical factor in enhancing accuracy than the choice of the "optimal" parametric distribution. We also find out that accounting for more explanatory factors mainly increases the resolution performances while it does not change the reliability contribution to the prediction performance metric considered (CRPS). 

\end{abstract}
	
\end{frontmatter}

\section{Introduction}

The prediction of upcoming atmospheric conditions is a very important
and challenging issue in a wide variety of domains. Accurate wind speed forecasting plays a pivotal role in various sectors, like transportation or disaster management. In particular, the short-term forecasting of surface wind speed that has to account for the highly intermittent nature of its fluctuations, is a difficult task 
that can be of prime interest for renewable energy production.  Prediction methods 
can be categorized using different criteria like the prediction horizon, the prediction objective
or the modeling framework \cite{Wang2021}. With regard to the latter, the most prominent approaches for predicting future atmospheric conditions rely on the so-called Numerical Weather Prediction (NWP) models, which, from the current state of the atmosphere, allow one to simulate its complex evolution in the next days.  Besides NWP models, that demand significant computational resources, wind speed forecasting techniques can rely on statistical methods, which analyze historical weather data to identify patterns and trends. Time-series analysis and models (like ARIMA  models) and machine learning algorithms are common statistical techniques applied in this context. These methods leverage historical wind speed observations to make predictions based on observed patterns and correlations. The methodology advocated in the article belongs to this overall framework of statistical and ``data driven" methods designed to provide accurate predictions at short-term .  

Let us note that a large majority of the methods that have been proposed so far in that field,
have been designed to produce {\em deterministic} or {\em point forecasts}, i.e., a single outcome that is obtained by minimizing a loss defined as, e.g., the mean absolute error or the mean square error. In the latter case,  the predicted value can be  interpreted as an estimation of the conditional expectation. We refer the reader to, e.g.,  \cite{Wang2021,7764085,hashli20,5619586,GIEBEL201759,kapisigi04} for a review on this topic.  In order to fully account for the uncertainty associated  with point forecast as well as for providing information about the likelihood of extreme events, probabilistic forecasting has emerged as a very active area over past few years. In principle, probabilistic forecasting enables decision-makers to mitigate the consequences of unpredictability and  make informed and optimal choices based on their risk aversion profiles. In the context of wind energy generation, reliable predictions on wind speed distribution allows for a better integration of this intermittent source in the power grid
\cite{Pinson2013}.  For this reason, many studies have been specifically devoted to the prediction of upcoming distribution of the wind speed itself or of the power of a wind turbine. Let us briefly describe those that are directly related to the present study, a complete review being beyond the its scope (see e.g.,  \cite{ReviewWindProbaForecasting2014,xie_overview_2023} for a recent review).

Historically, probabilistic forecasting in the general context of weather prediction involves "ensemble" forecasting methods.  Ensemble prediction mainly consists in generating multiple forecasts using different models, initial conditions, or parametrizations \cite{gneiting_Ensemble_05}. By aggregating these diverse predictions, ensemble methods offer more insight into the potential outcomes and associated uncertainties. Today, the major meteorological weather centres  routinely produce global ensemble forecasts, but in all these operational systems  the number of ensemble members is limited by the necessity to produce forecasts in a reasonable amount of time with the available computational power.  Alternatively or in complement to these ``Numerical Weather Prediction" (NWP) based estimation of the prediction uncertainty, many probabilistic forecasting methods relying on statistical models or Machine-Learning techniques have been proposed. Among these methods, the non-parametric ones are mainly quantiles-based approaches that involve the direct estimation of quantiles (such as, the median, the 90-th or the 10-th percentile) associated with the conditional  probability distribution \cite{Quantile_Regression_05,Nielsen_06}. The full shape of the forecast distribution can be then obtained by interpolating between the estimated quantiles \cite{HE2018374}. One advantage of quantiles-based approaches is that they do not require any assumptions about the underlying probability distribution of the forecast and are therefore adaptable to a wide range of forecasting situations. Another advantage is that the estimated quantiles can be used to calculate various uncertainty measures, such as exceedance probabilities or value-at-risk (VaR) estimates. Alternatively, parametric approaches involve the estimation of the parameters of a probability distribution that is assumed to underlie the forecast distribution.   The conditional distribution is then constructed using the estimated parameters. One advantage of parametric approaches is that they provide a more complete and yet more parsimonious characterization of the forecast distribution compared to quantiles-based approaches that involve more parameters to be estimated. Another advantage is that they provide a natural way to incorporate additional information, such as prior distributions or external data sources, into the forecast, using Bayesian framework.
One limitation of parametric log-likelihood approaches is that they can be sensitive to the choice of distributional assumption. In summary, quantiles-based approaches are more flexible and do not require any distributional assumptions, but they provide a less complete characterization of the forecast distribution.  

In this article, we focus on fully parametric approaches to probabilistic forecasting of surface wind speed for short term horizons. Our main goal is to compare the performances of different probability distributions already applied to the modeling of surface wind fluctuations due to their ability to  represent the conditional wind velocity probability density function. Our objective is also to quantify the benefits of using spatio-temporal information about wind distribution provided by neighboring stations as advocated in point forecast approaches (as in Ref. \cite{bamu_23} for instance).
One of the first papers addressing similar questions in the context of probabilistic forecasting is  Ref. \cite{gneiting_wind_speed_forecasting_2006} where the authors considered that the wind speed conditional law is a truncated normal distribution. A 2-hours ahead probabilistic prediction was then provided by regressing the conditional mean and variance parameters using a simple Regime Switching model that accounts for main wind directions and for past wind values including those of few upwind sites. They showed that their approach outperforms uni-variate time series models. Few years later, Pinson \cite{Pinson_2012} considered the same problem for very short term (i.e. few minutes ahead) wind power forecasting and proposed to model the conditional distribution of wind power as a generalized logit-Normal distribution which parameters were estimated using a recursive dynamic model.  In Ref. \cite{TastuPinson14}, the authors provided a probabilistic wind power forecasting for a single site of interest while using information from other wind farms as a dynamic explanatory variables in order to estimate the distribution parameters. They showed, for a set of wind farms in Denmark, that accounting for spatio-temporal effects improves the accuracy of probabilistic forecasts for prediction horizons ranging from few minutes to several hours. In Ref. \cite{holland_forecasting_2014}, a Weibull-based regression approach was considered  by minimizing 
the continuous ranked probability score (CRPS). This method appeared to perform better than both deterministic and probabilistic reference models on wind data collected from different sites in Japan.
In most of these early approaches, the forecasting step mainly consisted in regressing both location and scale parameters using a rather simple dynamical model.  In that respect they are easy to implement but they can struggle in capturing rich and complex dependencies between input factor and upcoming wind speed. 

Since then, more sophisticated Machine Learning models like Gaussian Process Regression and their extensions or Deep Learning methods \cite{XIANG2020113098,ZHU2019111772,Hossain21},  have proven to be of valuable interest in order  to estimate the parameters of the probability distribution of upcoming wind power or wind speed.  In Ref. \cite{Kou2013}, the authors used a so-called "warped Gaussian process" in order to predict the law of the wind-power, which they model as a non-linear transformation from the wind-speed itself considered as a Gaussian process.  In Ref. \cite{HU2017}, the authors built a model based on Student-t statistics for wind speed where the location parameter is a Gaussian process.  They showed that their model performs better that some already proposed  models designed for point prediction at horizons $h=30$ min and $h=1$ hour. In Ref. \cite{HENG2022118029}, the authors considered, a large variety of probability density functions (Weibull, Gamma, ....) for wind speed with a location parameter that is related to a latent Gaussian Process. This ``Generalized Gaussian Process'' framework allowed them to build different 6-hour-ahead probabilistic forecasts that they compared with each other and to existing linear models at several sites in China. To summarize, all these methods involve Gaussian processes in different variants as the main regression tool, notably for the location parameter of the condition law. They all provide interesting ideas and noticeable improvement as respect to some reference benchmarks. Let us however notice that none of them focused on the importance of surrounding sites as input factors even if one considers the benefits of using the past wind speed at few neighboring stations. Let us also emphasize that, as respect to Deep Neural networks, Gaussian processes are more computationally expensive, heavy to implement, less scalable and less capable of capturing complex relationships.   

While many recent publications concern Deep Learning applications to point forecasts or probabilistic forecasts by quantiles or intervals, there have been few studies exclusively devoted to parametric forecasting using deep learning techniques. One of the first and most famous approach in this topic is "DeepAR"\cite{DeepAR}, a recurrent neural network based methodology that can be trained on a large number of related time series in order to build efficient parametric probabilistic forecasts in different contexts.  In Ref. \cite{MASHLAKOV2021116405}, the authors proposed a review and an evaluation of this model and other similar global Deep Learning models, for probabilistic forecasting of energy production and notably wind energy generation. More recently, in Ref. \cite{ZHU2019111772} the authors combined a recurrent layer architecture (LSTM) with a Gaussian Process Regression in order to build a Gaussian posterior for wind speed probabilistic forecasting at very short term. In  Ref. \cite{Afrasiabi21},  a Deep Learning model made of a convolutional  block to handle spatial features and a recurrent layer to capture temporal dependencies, is proposed in order to predict the parameters of a Gaussian mixture representing the upcoming 1 hour ahead wind velocity. This paper undoubtedly presents an interesting approach, however the impact of neighboring spatio-temporal information on short-term forecasting performances is not considered.

In this work, we study, along the same line as DeepAR methodology, a Deep Neural Network model whose outputs are the parameters of a given wind speed probability distribution. As in \cite{DeepAR}, the log-likelihood function is used as the loss function to be optimized during the learning task. Within this powerful yet simple framework, our purpose is twofold. First of all, we want to compare different probability density functions (i.e. Truncated Normal law, Weibull, Gamma, Log-Normal,...) used classically to account for global wind speed fluctuations or as prior distributions in parametric prediction tasks as reviewed previously.  
We want to compare them  with each other but also with two Rice distribution based classes recently introduced to account for unconditional wind speed probability law: (i) the so-called ``Multifractal-Rice" (M-Rice) distribution \cite{BaMuPo10c,BaMuPo11} associated with the class of random cascade models that have proven to account remarkably well for the  fluctuations of surface wind speed \cite{BaMuPo10,BaMu10} and (ii) the "Rayleigh-Rice" (R-Rice) distribution that relies on a simple regime switching model of wind fluctuations introduced in \cite{drobinski_2015}. Secondly, another objective is to study in which respect one recovers the improvement in prediction accuracy when one accounts for spatio-temporal information as provided by the past wind speed values at neighboring locations.  To achieve these two goals, we consider a large variety of metrics or scoring rules that have been introduced in the context of probabilistic forecasting and that account for reliability and sharpness (or both) of the predicted probability density function. Our study relies on two hourly wind speed datasets recorded during the past two decades in two very different areas: in the Netherlands, which is a  flat territory in Northwestern Europe facing the North Sea, and in Corsica, an island in the French Mediterranean region characterized by a complex mountainous terrain. 

The paper is organized as follows : In section 2, we describe the two main data sets we employ within this study. Section 3 introduces various statistical tools and models that we consider.  It notably describes the  Deep Learning models we use and review various score and metrics useful to assess the quality of probabilistic forecasts. The main results are reported in  in section 4 while  section 5 contains conclusion and prospects for future research. Some technical material on CRPS decomposition are given in Appendix A while Appendices B,C  contain supplementary tables in order to provide complementary results.

\section{The data sets}
\label{sec:data}

\begin{figure}[t]
	\centering
	\includegraphics[width=0.95\columnwidth,angle=0]{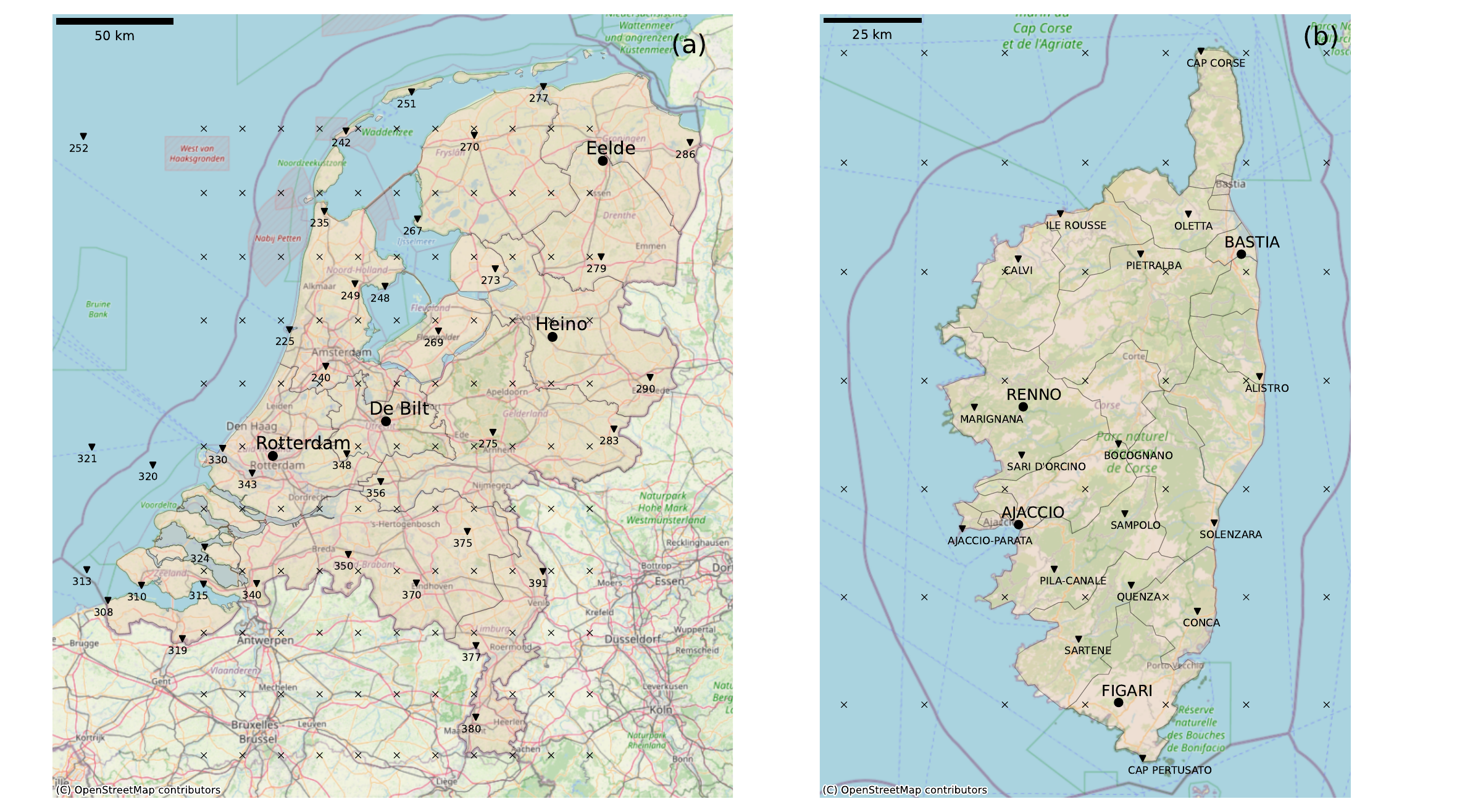}
	\caption{left map (a): Spatial distribution of the 42 stations over the Netherlands we use from the freely available Royal Netherlands Meteorological Institute wind series database. right map (b) : Locations of the 15 stations in Corsica from M\'et\'eo-France data. In (a) and (b),  ($\bullet$) symbols represent the stations where prediction is performed in the two regions, ($\triangledown$) the stations whose data are used in the prediction model and $ (\times$) the nodes of the ERA5 grid.}
	\label{fig_carte}
\end{figure}

\begin{table}
\resizebox{0.45\textwidth}{!}{%
	\begin{tabular}{c}%
		Corsica\\%
		
\begin{tabular}{|l|ccc|}
\hline
 site    &   $V_{max}(ms^{-1})$ &      
$k_{Weib.}$ &   $\sigma_{Weib.}$ \\
\hline
 Ajaccio &                22.0 & 1.98 &     3.53 \\
 Bastia  &                20.6 & 1.67 &     3.06 \\
 Renno   &                17.8 & 1.88 &     2.90 \\
 Figari  &                23.8 & 1.55 &     4.42 \\
\hline
\end{tabular}
\end{tabular}}
\resizebox{0.5\textwidth}{!}{%
	\begin{tabular}{c}%
		The  Netherlands\\%
\begin{tabular}{|l|ccc|}
\hline
 site    &   $V_{max}(ms^{-1})$ &      
$k_{Weib.}$ &   $\sigma_{Weib.}$ \\
\hline
 Heino     &                21.0 & 1.85 &     3.57 \\
 Eelde     &                22.0 & 1.92 &     4.73 \\
 De Bilt   &                17.0 & 2.00 &     3.84 \\
 Rotterdam &                24.0 & 1.84 &     4.97 \\
\hline
\end{tabular}
\end{tabular}}
\caption{Overview of the wind speed data in  the chosen sites, for the Corsica (left panel) and the Netherlands (right panel) datasets. Along with the maximum recorded velocity $V_{max}$, we illustrate the parameters $k_{Weib.}$ and $\sigma_{Weib.}$ of a fitted Weibull distribution (\ref{eq:wei}).}
\label{table_summary_dataset}
\end{table}

\begin{figure}[t]
	\centering
	\hspace*{-0.5cm}
	\includegraphics[width=0.95\columnwidth,angle=0]{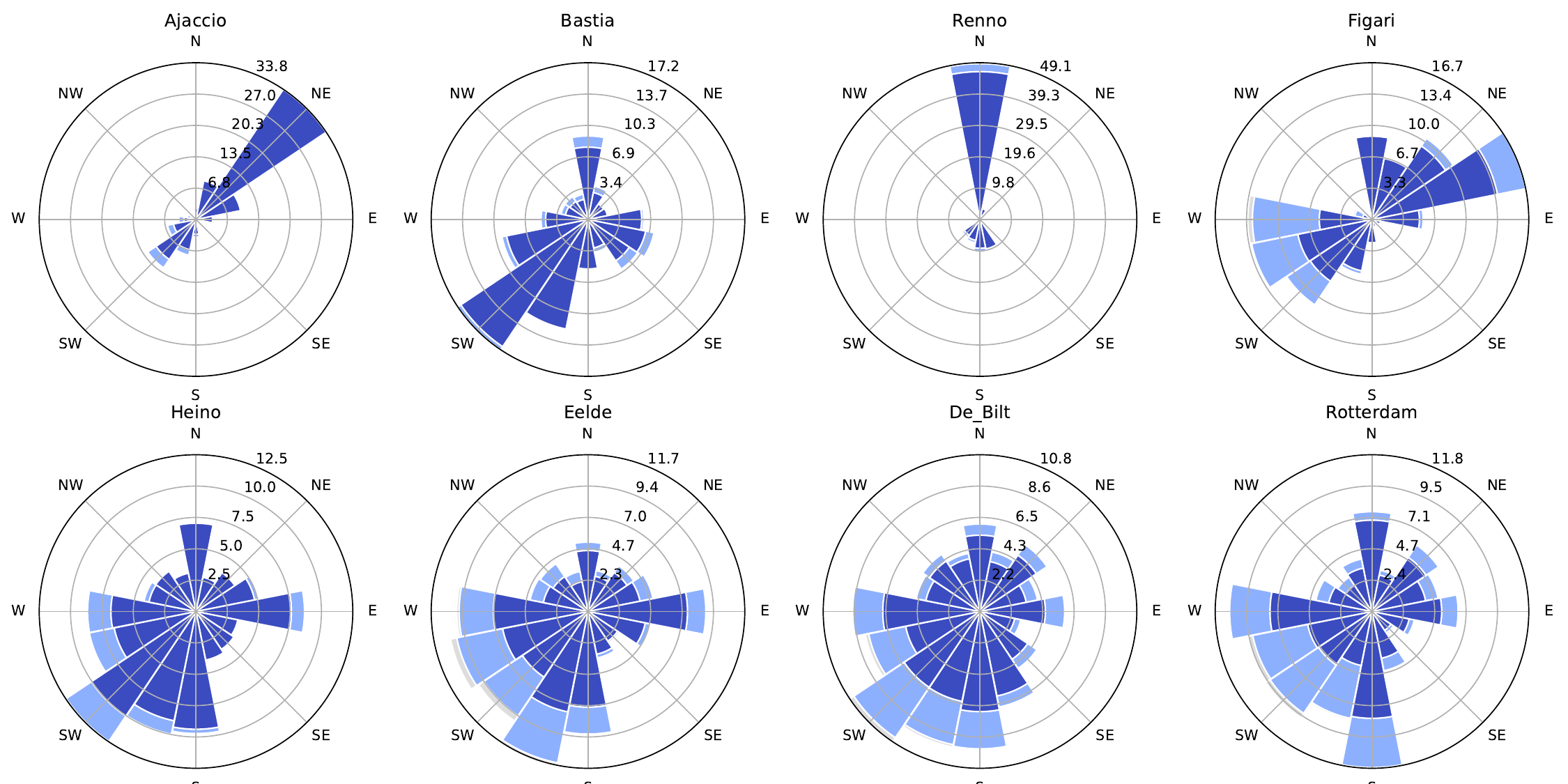}
	\caption{Wind roses of the 8 stations considered in Corsica (top row) and in the Netherlands (bottom row).}
	\label{fig_windrose}
\end{figure}

In this work, we use two different datasets of wind speed and direction time series. 
The first dataset is the same as the one considered in Ref. \cite{bamu_23}
for deterministic forecasting: it consists in a series of hourly mean wind speeds recorded by the Royal Netherlands Meteorological Institute at various stations spread over the Netherlands\footnote{KNMI time series records of weather data at various stations in the Netherlands are freely available online at https://www.knmi.nl/nederland-nu/klimatologie/uurgegevens.}.
The Netherlands is a flat and low-lying country in Northwestern Europe with an oceanic climate, 
featuring mild winters and cool summers and influenced by its proximity to the North Sea.
Hourly mean amplitudes and directions of the mean value during the 10-minute period preceding the time of observation of horizontal surface wind have been collected during very long time periods spanning more than 70 years for some stations. The locations of all the 42 sites we use as possible "neighboring" stations of a given site are reported in Fig. \ref{fig_carte} where ($\bullet$) symbols indicate the 4 ``reference'' stations where wind speed prediction is performed. The time period we consider for these stations extends from 01-01-2001 to 12-31-2020. The 42 stations were precisely chosen because they have few missing data during this time interval.  The  second dataset is quite different from the first one, as it consists in data at different locations in Corsica, a French Mediterranean island that mainly features mountainous landscapes with a Mediterranean climate characterized by hot, dry summers and mild, wet winters. The data have been collected, 
over the  period from 01-01-2011 to 12-31-2020, by M\'et\'eo-France\footnote{Corsica wind data have been provided by M\'et\'eo-France public Data service available at https://publitheque.meteo.fr/} at 20 different locations reported in Fig. \ref{fig_carte}(b). As in Fig. \ref{fig_carte}(a), the 4 sites we considered in the study are indicated by ($\bullet$) symbols while other station locations whose wind data are used as input, are indicated by ($\blacktriangledown$) symbols. For all the results described below, data randomly selected from the full dataset, representing a total size equivalent to 3 years (corresponding to 15 \% of the total KNMI data and 30\% of the MétéoFrance dataset), are used as the validation set to evaluate the model's performance, while the remaining data is used as the training set for optimizing the model parameters.

The 4 selected sites in each databases have a total number of missing data that is less than 15\% of the possible full data size, meaning that the size of the intercept of all periods where there is no missing data over all considered surrounding stations and considered past values, represents more than 85\% of the total possible number of data.
These stations are indicated ($\bullet$) in Fig. \ref{fig_carte}(a,b), and are located in Ajaccio, Bastia, Renno and Figari for the Corsica dataset, and Rotterdam, Heino, Eelde and De Bilt for data in the Netherlands.
 Notice that, as in \cite{bamu_23}, if there is a missing data during some given hourly period, we discard this time slice in the learning or in the validation sample. 
 
 As far as the main properties of wind speed distribution are concerned, it is worth mentioning that the wind speed resolution is $0.1 \; ms^{-1}$ for the M\'et\'eo-France data while it is only $1 \;  ms^{-1}$ in the case of KNMI data.   Table \ref{table_summary_dataset} summarizes the main features of the wind speed datasets in correspondence of the chosen locations by showing, along with the maximum module of wind speed recorded, the parameters of a Weibull distribution (\ref{eq:wei}) fitted to the station time series.   We can notice that the Weibull shapes are more homogeneously close to the Rayleigh distribution ($k=2$) in the Netherlands than in Corsica. This spatially less  homogeneous nature of the wind distribution can also be observed  in Fig. \ref{fig_windrose} where are displayed the wind roses of all 8 stations. One can first see that, for all 4 stations in the Netherlands, the prevailing wind regimes are very similar and all come from South-West. In Corsica the situation is very different, as a consequence of the specific geographical situation and the complex orography of the island  pattern.

Let us mention that, in some of the experiments of our study, we also make use of re-analysis data as input of our probabilistic forecasting model. For that purpose we use ERA5 data. ERA5 is the acronym for the 5-th generation ECMWF reanalysis for the global climate and weather \cite{ERA5}. It provides hourly estimates for a large number of atmospheric, ocean-wave and land-surface quantities on a regular latitude-longitude grid of mesh size 0.25 degree. The ERA5 grid points are indicated by symbols ($\times$) in Figs. \ref{fig_carte}(a,b). In this study, we restrict ourselves to the use of eastward and northward 10 m wind speed components $(u_{10},v_{10})$ at the $15$ grid points closest the considered station.

Finally, it is important to note that all the anemometers at the considered stations for both datasets are situated 10 meters above ground that is significantly below the relevant heights for the purpose of energy production. Despite the acknowledged differences between surface winds and hub-height winds, we believe that the approach advocated in this work remains  interesting in wind energy applications if one assumes, in line with the practices of 
numerous prior studies in the field (see, e.g., \cite{Grassi2015,Amato22,Lopez2022,IEA23}), that one can employ conversion methods to extrapolate data from 10 $m$ to higher heights.
Nevertheless, it would be important to confirm the accuracy of such practices and perform specific studies relying on actual wind values at heights equal or above 100 meters \cite{Drechsel12}. Unfortunately, comprehensive databases containing such time series are not readily available 
and, as mentionned in Sec. \ref{s_conc}, such an analysis is left for future research.

\section{Prediction method, probability densities, probabilistic losses and metrics}
\label{sec:reg}

\subsection{Deep Neural Network models for parametric probabilistic prediction}
\label{sec:models}
As discussed in the introduction, a primary goal of this paper is to evaluate different potential forms of the upcoming wind probability density function (pdf) 
for the purpose of conducting parametric probabilistic forecasting within short-term horizons.
Before describing precisely our approach, let us first introduce some useful notations. 
Let $V^s_t$ ($V^s_t > 0$) and $\varphi^s_t$ ($0 \leq \varphi^s_t < 2 \pi$) stand  respectively for the amplitude and the direction of the hourly mean wind velocity at time $t$ and for station $s$. 
We denote by $u^s_t$ and $v^s_t$ the Cartesian (i.e. eastward or northward) components of the surface wind velocity at station $s$, namely, $u^s_t = V^s_t \cos(\frac{3 \pi}{4} - \varphi^s_t)$ and $v^s_t = V^s_t \sin(\frac{3 \pi}{4}-\varphi^s_t)$ \footnote{Let us mention that the transformation $\varphi \rightarrow\frac{3 \pi}{4}-\varphi$ results from the fact that the reported wind direction corresponds to the direction from where the wind is coming with the convention that $N \equiv 360^o$ and $E \equiv 90^o$.} and then by $\bu_t$ and $\bv_t$ the corresponding (N+1)-dimensional vectors associated with all stations considered in the experiment (that is, the reference site ($\bullet$)  along with N neighbors chosen among the available stations indicated by symbols ($\blacktriangledown$) in Fig. \ref{fig_carte}). In order to account for seasonal (i.e. annual) and also for diurnal (i.e. daily) variations in wind regimes,  we use the same convention as in  Ref. \cite{bamu_23}: let $hh_t \in [1, \ldots, 24]$ and $dd_t \in [1, \ldots, 365]$ denote respectively the hour of the day and the day of the year associated with time $t$ and define $\bS_t$ as the 4-dimensional vector:
\be
\label{eq:defS}
\bS_t = \left.
\begin{bmatrix}
	\cos(\frac{2 \pi hh_t}{24}) \\
	\sin(\frac{2 \pi hh_t}{24})  \\
	\cos(\frac{2 \pi dd_t}{365}) \\
	\sin(\frac{2 \pi dd_t}{365})  
\end{bmatrix}
\right. \; .
\en

In order to build the model input of our experiments, let us define the  $N_f$-dimensional vector with $N_f = 2(N+1)+4$:
$$
\bX_{t}= \left. 
\begin{bmatrix}
	\bu_t \\ \bv_t \\ \bS_t 
\end{bmatrix}	
\right. \,.
$$

The features tensor given in input to the model is then the matrix formed by vectors $\bX_t$, with time varying from $t$ up to $N_{t}-1$ steps in the past ($N_t \geq 1$).
It is therefore a $N_{f}\times N_t$  matrix of the type:
\begin{equation}
\label{def:XtN}
	 \bX_{t,N_{t}}  = \left[ \bX_{t-N_t+1}, \bX_{t-N_t+2} , \ldots , \bX_{t} \right]\; .
\end{equation} 
\begin{figure}[t]
	\centering
	\hspace*{-0.5cm}
	\includegraphics[width=0.7\columnwidth]{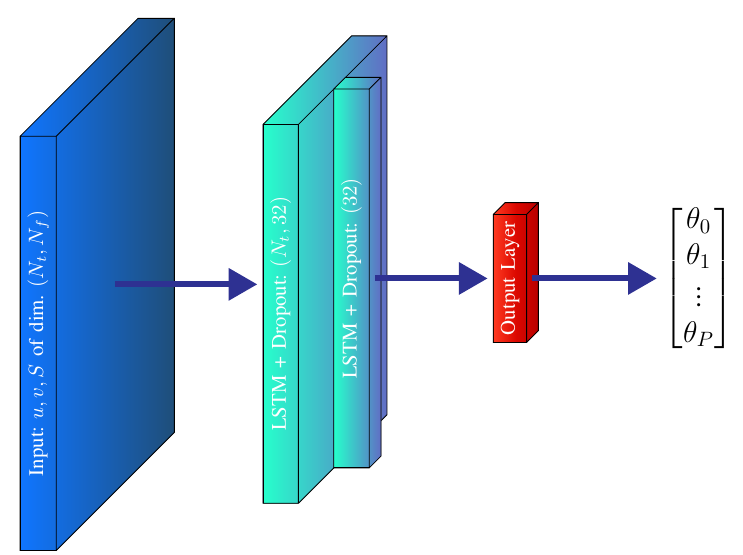}
	\caption{Sketch of the LSTM network model used for wind distribution parameter prediction. The depth of each block symbolizes the size of past time data. The green blocks represent the LSTM layers. The first one returns all intermediate states while the second one returns only the last state. The red block represents the output layers that is mainly composed of a single fully connected layer with appropriate individual activation function layers.}
	\label{fig_lstm}
\end{figure}
Our objective is to define a model $\mathcal{M}$ that takes as input the tensor of explanatory variables $\bX_{t,N_{t}}$ and 
that returns the $P$-dimensional parameter vector $\bT_{t+h}$ of $f(V |\bT)$,  a  probability density function representing  
the conditional wind speed distribution at horizon $t+h$:
\begin{equation}
  \bT_{t+h} = \mathcal{M}(\bX_{t,N_{t}})
\end{equation}

The class of non-linear models $\mathcal{M}(\bX_{t,N_{t}})$ we consider in this paper are parametrized by some specific
artificial neural network (ANN) models.  ANN can be mainly defined as the composition of $L$ intermediate non-linear functions (the first one being referred to as the `input layer", the other ones are called ``hidden layers'') and a final output function (the ``output layer''). One generally speaks of ``Deep Neural Network'' (DNN) when the number of hidden layers is large enough, say $L \geq 2$ (see e.g., \cite{Good16}). DNN are at the heart of a large number of studies devoted to wind speed forecasting and have proven to be among the most efficient methods.  In the case of probabilistic wind speed forecasting, various DNN models have been proposed, mainly for quantiles or intervals evaluation \cite{Wu_2016,Afrasiabi21,xie_overview_2023}.  There is a huge variety of neural networks and in this work our ambition is not to propose a new model or to find the ``best'' one among the vast choice of solutions that have already been explored. We rather chose to follow the approach proposed in ref. \cite{DeepAR}, based on recurrent neural network and that has proven to perform very well for a wide range of applications.
The precise model we consider is depicted in Fig. \ref{fig_lstm} where 2 LSTM layers each with 32 units  are stacked in front of the output layer.
This later directly provides the probability law parameters up to a final non-linear function (namely, as explained in the next section, an exponential, a softplus or a sigmoid transformation) that accounts for their expected range of values. We have compared the performances of this architecture with other alternatives relying on fully dense layers, convolution layer or even the recently considered attention layers for time-series \cite{vaswani2017attention}. All these considered models provide comparable results with a small advantage to the LSTM network that we chose to present. 
Besides these rather classic neural  network models, we also have considered, 
as a reference model, the approach pioneered in \cite{gneiting_wind_speed_forecasting_2006} where the authors proposed a so-called ``Regime Switching Space-Time" (RST) method to 
predict the probability density function of the wind speed at a two hours time horizon. 
More specifically, we extend the RST model in order to account for an arbitrary number of neighboring stations and to encompass a broader range of wind regimes. In Appendix \ref{sec:app_4}, we
describe in details this model and a fully connected alternative to the LSTM network introduced above. We also present the key performance metrics of both models  for comparative analysis.

 As far as the loss function is concerned, as in \cite{DeepAR}, the estimated density parameters are those that minimize the negative log-likelihood, (also referred to as the \emph{logarithmic score} in the case of performance evaluation as detailed later on) which is simply defined as: 
\begin{equation}
\label{def:logS}
   \ell(y_{t},\bT) = - \ln f(y_{t}|\bT)
\end{equation}
where $f(y_{t} | \bT)$ stands for the probability density function of parameter $\bT$ and evaluated at ``label'' $y_{t}$, namely the amplitude of the wind speed at the site of interest $s$ and at forecasting horizon $h$: $y_t=V^s_{t+h}$.  In the next section we describe the various families of the wind speed pdf we consider in this study.

\subsection{Wind speed probability density functions}
\label{sec:dist}
One of the critical issues in parametric probabilistic forecasting is the choice of the family
of probability distribution one uses. In the case of surface wind speed, several laws have been proposed in the literature,
whether for modeling the overall wind distribution or for the purpose of probabilistic forecasting.
The simplest law, one can consider for the wind speed probability density function
is undoutbedly a simple Gaussian law. More precisely, since only positive values
can occur, some authors consider the so-called truncated normal distribution 
which is defined as the restriction of the normal law to the positive real axis. 
Its density is therefore:
\begin{equation}
	  f_{tN}(y | \mu, \sigma) = \frac{e^{-\frac{(y-\mu)^2}{2\sigma^2}}}{  Q(-\frac{\mu}{\sigma}) \sqrt{2 \pi \sigma^2}}
	\label{eq:tNorm}
\end{equation}
where $\mu$ and $\sigma^2$ are the mean and variance parameters of the corresponding Gaussian law and 
$
Q(z) = \frac{1}{\sqrt{2 \pi}} \int_{z}^{\infty} e^{-\frac{u^2}{2}} du = \frac{1}{2} \mbox{erfc}(\frac{z}{\sqrt 2})
$ is the complementary cumulative distribution function of the standard normal law. Let us mention that this distribution is notably used in the regime switching  model proposed in \cite{gneiting_wind_speed_forecasting_2006} which is one the first proposed statistical 
model aiming at probabilistic short-term prediction  of surface wind speed.

Among the laws used to describe wind speed distribution, the most widely used is the Weibull distribution \cite{review_wind_speed_distributions_09,review_wind_Speed_distributions_21} whose expression reads:
\begin{equation}
	 f_{W}(y | \; k,\sigma) = \frac{k}{\sigma} \left( \frac{y}{\sigma}\right)^{k-1} e^{-(\frac{y}{\sigma})^k}
	 \label{eq:wei}
 \end{equation}
where $\sigma$ is a scale parameter and $k$ a shape parameter. Let us notice that when $k=1$ the Weibull distribution reduces to an exponential distribution while when $k=2$, this distribution corresponds to the so-called Rayleigh law. The Weibull parameters measured by fitting the historical wind variations at a given site are commonly used to quantify the wind power potential of this location \cite{conradsen}. The log-normal law corresponds to a random variable which logarithm is a Gaussian (normal) random variable.  It has often been considered as an alternative to the Weibull distribution in the context of wind power potential evaluation \cite{review_wind_speed_distributions_09}.  If $\mu$ and $\sigma$ stand respectively for the mean and the scale of the random variable logarithm, the expression of its log-normal law reads:
\begin{equation}
	f_{L}(y | \mu,\sigma) = \frac{1}{\sqrt{2 \pi} y \sigma} e^{\frac{(\ln y - \mu)^2}{2 \sigma^2}} \; .
\end{equation} 

Another class of distributions that is often considered in the context of wind energy is the Gamma distribution \cite{review_wind_Speed_distributions_21}  that is also a two-parameters family that generalizes the exponential law. It is defined as:
\begin{equation}
	f_{G}(y | \; k,\sigma) = \frac{1}{\Gamma(k) \sigma} \left( \frac{y}{\sigma}\right)^{k-1} e^{-\frac{y}{\sigma}}
	\label{eq:gamma}
\end{equation}
where again $\sigma$ and $k$ are respectively the scale and the shape parameters while $\Gamma(z)= \int_0^\infty t^{z-1} e^{-t}  dt $ stands for the standard Gamma function.

Another distribution, related to (\ref{eq:gamma}), that has been considered more recently and that provides appealing results in the field wind speed fluctuations is the Nakagami distribution \cite{Dookie18,yu19,Gugliani2020,suriadi21}, whose expression, for a shape parameter $m$ and a scale parameter $\sigma$, reads:
\begin{equation}
	f_{N}(y |m,\sigma) = 2 \frac{ m^m}{y \Gamma(m)}  \left(\frac{y}{\sigma} \right)^{2m} e^{-m \frac{y^2}{\sigma^2}} \; .
\end{equation} 

Let us end this short review of the different probability laws we consider, by introducing 3 ``Rician'' probability distributions. Let us recall that the 'Rice' distribution  \cite{Rice} of centering parameter $\nu$ and scale $\sigma$ corresponds to the law of the norm (or modulus) of a 2-D random vector whose components are two independent Gaussian random variables of same variance $\sigma^2$ and with a mean $\nu_1$ and $\nu_2$ such that $\nu = \sqrt{\nu_1^2+\nu_2^2}$. 
In other words, $\nu$ corresponds to the length of the vector built from the two mean values while $\sigma^2$ is the variance its components.
The corresponding pdf can be written as follows:
\begin{equation}
	f_{R}(y | \nu, \sigma ) =  \frac{y}{\sigma^2}e^{-\frac{y^2+\nu^2}{2 \sigma^2}} I_0(\frac{y\nu}{\sigma^2})
\end{equation} 
where $I_0(y)$ is the modified Bessel function of the first kind with order $0$. Notice that when $\nu = 0$, one recovers the expression of the Rayleigh distribution. The Rice distribution has been considered directly as a model for wind speed distribution (see e.g., \cite{Gugliani2020}). Recently, however, two different 3-parameters distributions, whose construction is based on the Rice distribution, have been proposed and they appear to provide very good performances to fit surface wind speed empirical distributions. One of  the advantages of Rice-based distributions  is that they can rely on random Gaussian fluctuations for the $(u,v)$ velocity components that result from a simple dynamical  (physical or empirical) model \cite{BaMuPo10c,drobinski_2015}. Moreover,  under relatively mild conditions, these laws remains stable as respect to conditioning by a set of observations and in that respect they can  account for both conditional and unconditional (i.e., the "climatology") wind speed fluctuations.
The ``Rayleigh–Rice distribution'' has been introduced in \cite{drobinski_2015} in order to account for wind anisotropy and the existence of different wind regimes.  In its simplest version, this distribution is defined as a mixture of two Rice distributions with the same scale parameter $\sigma$, namely the Rayleigh distribution standing for a
``random flow'' with no preferential direction and a  ``channeled flow" where  $(u,v)$ have different, non-zero, mean values. Its expression then reads:
\begin{eqnarray}
	f_{RR}(y | \alpha,\nu, \sigma )  &= &  (1-\alpha) f_{R} (y,0,\sigma)+ \alpha f_R(y | \nu,\sigma) \\
	                                                    &=&  (1-\alpha)   \frac{y}{\sigma^2} e^{-\frac{y^2}{2 \sigma^2}} + \alpha 
	                                                      \frac{y}{\sigma^2} e^{-\frac{y^2+\nu^2}{2 \sigma^2}} I_0(\frac{y\nu}{\sigma^2})
\end{eqnarray} 
where the mixture parameter, $0 \leq \alpha \leq 1$, simply corresponds to the probability of being in the ``channeled regime'' \cite{drobinski_2015}.  The Rice distribution has also been considered in ref. \cite{BaMuPo10c} where the authors introduced the so-called ``Multifractal-Rice'' (M-Rice) distribution that consists in considering a Rice-distribution whith a random scale parameter
that follows, e.g., a log-normal distribution. This model corresponds to a model of stochastic dynamics where the fluctuations of $(u,v)$ are described by a random cascade as introduced in the context of fully developed turbulence.
In this framework, the M-Rice distribution can be written as \cite{BaMuPo10c}:
\begin{equation}
	f_{MR}(y | \nu, \sigma, \lambda^2 )  =    \int P(\omega) f_{R} (y,\nu, \sigma e^{\omega}) \; d \omega 
\end{equation}
where $P(\omega)$ represents the centered law of the scale logarithm. As in the turbulence literature, this law is chosen to be a Gaussian law of variance $\lambda^2$ which yields\footnote{In turbulence the parameter $\lambda^2$ is referred to as the ``intermittency coefficient'' \cite{Fri95}.}:
\begin{eqnarray}
	f_{MR}(y | \nu, \sigma, \lambda^2 )	&=&  \frac{1}{\sqrt{2 \pi \lambda^2}}  \int e^{-\frac{\omega^2}{2 \lambda^2}} f_{R} (y,\nu, \sigma e^{\omega}) \; d \omega  \\
	&=&   \frac{1}{\sqrt{2 \pi \lambda^2}}  \int e^{-\frac{\omega^2}{2 \lambda^2}}   \frac{y}{e^{2\omega}\sigma^2} e^{-\frac{y^2+\nu^2}{2 e^{2 \omega} \sigma^2}} I_0(\frac{y\nu} {e^{2 \omega} \sigma^2}) \; d \omega 
\end{eqnarray}

Let us notice that the last formula can evaluated numerically using a Gauss-Hermite quadrature \cite{abst64}:
\begin{equation}
  \int _{-\infty }^{+\infty }e^{-y^{2}}f(y)\,dy \approx \sum _{i=1}^{n}w_{i}f(y_{i})
\end{equation}
were $n \geq 1$, the abscissa $\{y_i\}_{i=1,\ldots,n}$ correspond to the roots of the order $n$-th Hermite polynomial $H_n(y)$ and the weights $\{w_i\}_{i=1,\ldots,n}$ are:  
$$
  w_{i}={\frac {2^{n-1}n!{\sqrt {\pi }}}{n^{2}[H_{n-1}(y_{i})]^{2}}}
$$
In practice, we observe that $n=7$ is sufficient for the purpose of this paper and we have checked that increasing the quadrature order up to $n=11$ does not bring any noticeable  improvement.

Let us notice that all the distributions involve only 2 parameters whereas Rayleigh-Rice and M-Rice are characterized by 3 parameters.  In order to compare all these distributions, one could try,  as for the Akaike Information Criterion (AIC) or the Bayesian Information Criterion (BIC), to incorporate a penalty that accounts for the number of parameters in the log-likelihood or in other scores. But this makes sense only if one evaluates the model on the same data that was used for the fitting, namely on the training period. Indeed, the penalties in AIC or BIC are designed to offset the bias caused by overfitting when both evaluation and fitting are conducted on the same dataset. When the models are evaluated on test data that are different from the training data, there is no such a bias and therefore there is no need of any penalty. 

Let us end this review by mentionning that, if $x_1,x_2, \ldots, .$ stand for the outputs of the last layer of our Neural network, they do not directly correspond to the law parameter we want to forecast. The variance parameter of the considered law is always computed as $\sigma = e^{x_2}$ while the shape parameter $k$ in the case of Weibull, Gamma, $m$ for Nagakami or the mean parameter $\nu$ for Rician distributions is built from the softplus function, i.e., $k,m,\nu = \ln(1+e^{x_1})$. In the case of a truncated-normal or of a log-normal law, one takes directly $\mu = x_1$. Finally, since they are between $0$ and $1$, the parameters $\alpha$ and $\lambda^2$ for Rayleigh-Rice and M-Rice distributions respectively are obtained from $x_3$ using the logistic function: $\alpha,\lambda^2 = \frac{1}{1+e^{-x_3}}$.

\subsection{Scoring rules for the evaluation of probabilistic forecasts}
\label{sec:metrics}
Assessing the performances of a given forecasting model is much more challenging in the case of probabilistic forecasting than in the case of point forecasts. For that reason, the study and the  definition of suitable skill measures for probabilistic  forecasts is still an active area of research \cite{gneiting2014probabilistic,du2021beyond}. In this section, we present briefly the scalar scoring rules which we use throughout this work to assess the performance and to compare the different forecast models we consider.

Let $f_k$ be the forecast probabilistic distribution and $y_k$ the corresponding realization of the observed random variable $Y$. A scoring rule $S(f_k,y_k)$ assigns a scalar value to the probabilistic forecast $f_k$ on the basis of the forecast-observation pair $(f_k,y_k)$. All the scoring rules we consider are negatively oriented, that is, the smaller their value the more accurate is the forecast.
Moreover, most of the considered scores are \emph{strictly proper}. When speaking of probabilistic forecasts, a score is said to be proper if its expected value $\mathbb{E}_{g}S(f,Y)$ is minimized when $f=g$ (being $g$ the "true" probability distribution of the observations $Y$), and \emph{strictly proper} if such minimum is unique \cite{gneiting2007strictly,gneiting2014probabilistic}. When a scoring rule it is not proper, it means that scores can be improved artificially by \emph{hedging} \cite{jolliffe2008impenetrable}, that is, the best score may correspond to a forecast that does not match the true forecaster's beliefs. Therefore, it is advisable to use scoring rules which are proper to avoid misguided interpretations of the results \cite{gneiting2007strictly}.

In practice, forecasts are compared by taking the average score on the $k=1,\ldots,M$ available realizations $(f_k,y_k)$. In the following, we add a $\cdot _k$ subscript to the name of the score when it refers to a single $(f_k,y_k)$ pair, and otherwise assume it has been averaged on all the $M$ available instances (times).

Among scoring rule which are strictly proper, one of the most used is the \emph{logarithmic score} (LogS) \cite{good1952rational}:
\begin{equation}
 \text{LogS}_k = \text{LogS} (f_k,y_k)=-\ln{f_k(y_k)}\,.
	\label{logs}
\end{equation}
This score is sometimes called \emph{ignorance score}, with reference to its relation to Shannon information content \cite{roulston2002evaluating,gneiting2007strictly,diks2011likelihood} and is also referred to as the negative log-likelihood $\ell(y_{k},f_k)$ as discussed before Eq. \eqref{def:logS} . In the context of this work, the focus on this score is natural as we use the negative log-likelihood to train the DNNs which give the parameters of the conditional distributions $f_k$. We remark that in the literature the use of the logarithmic score is not always recommended, as some authors consider the penalties associated with extremes  too high \cite{selten1998axiomatic}, while others see as a limitation the fact that it is a \emph{local score}, and suggest the use of non-local scores instead \cite{gneiting2007strictly,stael1970family}.  However, there seems to be a lack of consensus about the desirability of non-locality \cite{du2021beyond}.
A widely used nonlocal score which is also strictly proper is the \emph{continuous ranked probability score} (CRPS) \cite{brown1974admissible,hersbach2000decomposition}, defined as: 
\begin{equation}
	\text{CRPS}_k = \text{CRPS} \left( f_k,y_k\right)=\int_{0}^{\infty}\left( F_k(y) - H(y - y_k)\right)^2 dy\,,
	\label{crps}
\end{equation}
where $F_k(z) = \int_{-\infty}^z f_k(x) dx $ is the Cumulative Density Function (CDF) of $f_k$ and $H(y - y_k)$ is the Heaviside step function centered at observation $y_k$. In the case of a deterministic forecasts, $f_k(y)$ is a Dirac delta function and (\ref{crps}) reduces to the expression of the absolute error, its average expression on $M$ samples reduces to the \emph{Mean Absloute Error} (MAE) which gives this score a natural interpretation \cite{hersbach2000decomposition}. A non-local score analogous to \eqref{crps}, but based on the logarithmic score, the so-called  \emph{Continuous Ranked IGNorance score} (CRIGN) has been proposed in \cite{todter2012generalization}, but its use is less common in the literature. Notice that the score (\ref{crps}) has a dimension (the same of the random variable considered). This implies that values of CRPS are not "absolute penalties" but should always be understood as relative to a certain unit measure ($ms^{-1}$ in the tables of Sec. \ref{sec:results} below).   

Proper scoring rules account for both \emph{calibration} and \emph{sharpness} \cite{gneiting2007probabilistic}. Calibration refers to the statistical consistency between forecasts and observations and it is a property of the forecast/observation pair. Sharpness is a property of the forecast distribution alone, and refers to the  degree of dispersion of the distribution. The sharper the distribution, the less spread the distribution and the lower the forecast uncertainty. Naturally, sharpness is a desired quality only when the considered forecast is calibrated. At the same time, once calibration has been ensured, sharpness plays an important role in discriminating between forecasts that contain conditionally more information with respect to the climatology distribution \cite{gneiting2007probabilistic}. It has been shown that all strictly proper scores admit a decomposition in a reliability, a resolution and a uncertainty term \cite{brocker2009reliability}. The first and the second terms are related to calibration and sharpness respectively, while the latter is related to the available observation and is independent from the forecasts. The smaller the reliability component the more ``trustworthy" the forecast, while the higher the resolution the better the forecast. In the Tables of \ref{sec:app_2}, we provide the values of the relative contributions of the reliability and the resolution components of the CRPS, based on the decomposition proposed by Hersbach in \cite{hersbach2000decomposition}. Such a decomposition is described 
in the context of a continuous distribution in \ref{sec:app_1}). 

For the sake of completeness, let us discuss two more scores measuring respectively calibration and sharpness. In the literature, calibration is often evaluated by means of the \emph{Probability Integral Transform} (PIT) \cite{diebold1997evaluating}:
\begin{equation}
	z_k=\int_{0 }^{y_k}f_k\left( y\right)dx=F_k(y_k) \; .
	\label{pit}
\end{equation}
For the probabilistic forecasts to be calibrated, the $k=1,...,M$ transforms $z_k$ should have a standard uniform distribution $U(0,1)$, in the limit of sample variability. Distribution of the $z_k$ is usually displayed visually by means of a \emph{PIT histogram}.  The information provided by the PIT histogram can be summarized using the \emph{reliability index} (RI) \cite{delle2006probabilistic}, which accounts for the absolute distance of the $z_k$ from uniformity:
\begin{equation}
	\text{RI}=\frac{1}{M}\sum_{j=1}^{N_{bs}}\left| n_j - \frac{M}{N_{bs}} \right|\,,
	\label{ri}
\end{equation}
where $N_{bs}$ is the total number of histogram bins while $n_j$ is the number of counts in the $j$-th histogram bin. Obviously, the PIT histogram and the associate RI depend on the chosen number of bins $N_{bs}$. In practice, $10$ to $20$ bins are considered as sufficient.
Following \cite{pinson2007non}, an additional measure of sharpness can be constructed using the average size of the confidence interval at level $\beta$ associated with the predictive distributions $f_k$: 
\begin{equation}
	\text{Sharp}=\frac{1}{M}\sum_{k=1}^M \left( q_k(1-\beta/2)-q_k(\beta/2)\right) \,,
	\label{sharpness}
\end{equation}
where the $q_k(\theta)$ stands for the quantiles at level $\theta$ of the law $f_k$. In this paper we choose $\beta=0.8$, but, as in \cite{pinson2007non}, one can also consider a sharpness measure for a full range of quantile values.
As mentioned above, a good calibration or a good sharpness are not, if considered alone, indicative of a good forecast. Indeed, the climatology distribution is perfectly calibrated, but not very informative, and sharp distributions are desirable only when calibrated. Therefore RI (\ref{ri}) and Sharp (\ref{sharpness}) add additional information when comparing forecasts, but the strictly proper scores LogS (\ref{logs}) and CRPS (\ref{crps}) are more informative if considered alone. 

While the scores above consider the overall performance of a forecast model, another important property is the capability of predicting extreme events.  Even if the focus of this study is not on the quantifying the likelihood of the occurrence of extreme winds, one would like to assess the forecasting performances of our models in situations when very strong winds are observed. For that purpose, we consider as ``extreme'' all the events whose value lie beyond a large threshold quantile $q^c_{\beta^\star}$ of the climatology distribution, i.e., $1-\beta^\star \ll 1$ (in practice we choose $\beta^\star = 0.95$). 
The choice of a suitable score is not straightforward. While it may be tempting to use (\ref{logs}) and (\ref{crps}) with restriction to the observations matching the definition of extreme event, the so obtained scores are improper and favor forecast models biased towards extremes  (see \cite{lerch2017forecaster}). A strictly proper version of the CRPS which focuses on extreme events is the \emph{threshold-weighted continuous ranked probability score} introduced by \cite{ranjan2010combining}:
\begin{equation}
	\text{twCRPS}_k  = \text{twCRPS} \left( f_k,y_k\right)=\int_{0}^{\infty}w(y)\Big( F_k(y) - H(y - y_k)\Big)^2 dy\,,
	\label{twcrps}
\end{equation}
with weight function $w(y)=\mathbb{I}(y\geq q^c_{\beta^\star})$. 
Another strictly proper score which has been designed especially for evaluating forecasts performances at extremes is the \emph{censored likelihood} (CSL) \cite{diks2011likelihood}:
\begin{equation}
	\text{CSL}_k  = \text{CSL} \left( f_k,y_k\right)= -w(y_k)\log f_k(y_k)- 
	\left(1- w(y_k) \right)\log\left(1-\int_{0}^{\infty}w(y)f_k(y)dy\right)\,.
	\label{csl}
\end{equation}
As before, the ranking of competing forecast models is made by the average on all $i=1,..,M$ available forecast-observation pairs, that is $\text{twCRPS}=\frac{1}{M}\sum_k \text{twCRPS}_k$  and $\text{CSL}=\frac{1}{M}\sum_k \text{CSL}_k$.

\subsection{Point forecast evaluation metrics and reference models}
\label{sec:pforecast}

Let us end this section by noticing that a probabilistic prediction can also be used to compute a point forecast $\hat{y}_k$ and can therefore be directly compared to other models for single predictions. For instance in order to minimize the Root Mean Square Error (RMSE), 
\begin{equation}
	RMSE = \sqrt{\frac{1}{M} \sum_{k=1}^M \big(\hat{y}_k - y_k \big)^2}
\end{equation}
it is optimal to compute the conditional mean, i.e.,  
\begin{equation}
\label{eq:cmean}
	\hat{y}_k = m_k = \E \big[{y_k} \big] = \int_0^\infty f_k(y) y dy \; ,
\end{equation}
where we have used the same notations as in previous section. 
If one wants to minimize the Mean Absolute Error (MAE):
\begin{equation}
 MAE =  \frac{1}{M} \sum_{k=1}^M \big| \hat{y}_k - y_k \big|
\end{equation}
it is better to estimate the conditional median value:
\begin{equation}
\label{eq:cmed}
	 \hat{y}_k = q_k({0.5}) = F_k^{-1}(0.5)
\end{equation} 
where $F_k^{-1}$ is so-called percent-point function, the inverse the cumulative distribution function at step $k$.
Within this "single-point" framework, two classic benchmarks can be considered, namely the so-called  {\em persistence} model and the class of linear models. The persistence model simply consist in assuming that the best predicted wind speed is the last observed one while a linear models obtain the predicted wind speed value 
as a linear regression from input features, the coefficients of the regression being usually obtained as the ones that minimize the Mean Square Error.  Such an approach encompasses classic time-series models like ARMA or Vector Auto Regressive (VAR) models.  We refer the reader to e.g., Ref. \cite{bamu_23} for a comparison of these two basic models to more complex Machine Learning  approaches. 


\section{Empirical results}
\label{sec:results}
\begin{figure}[t]%
	\centering%
	\includegraphics[width=0.8\linewidth]{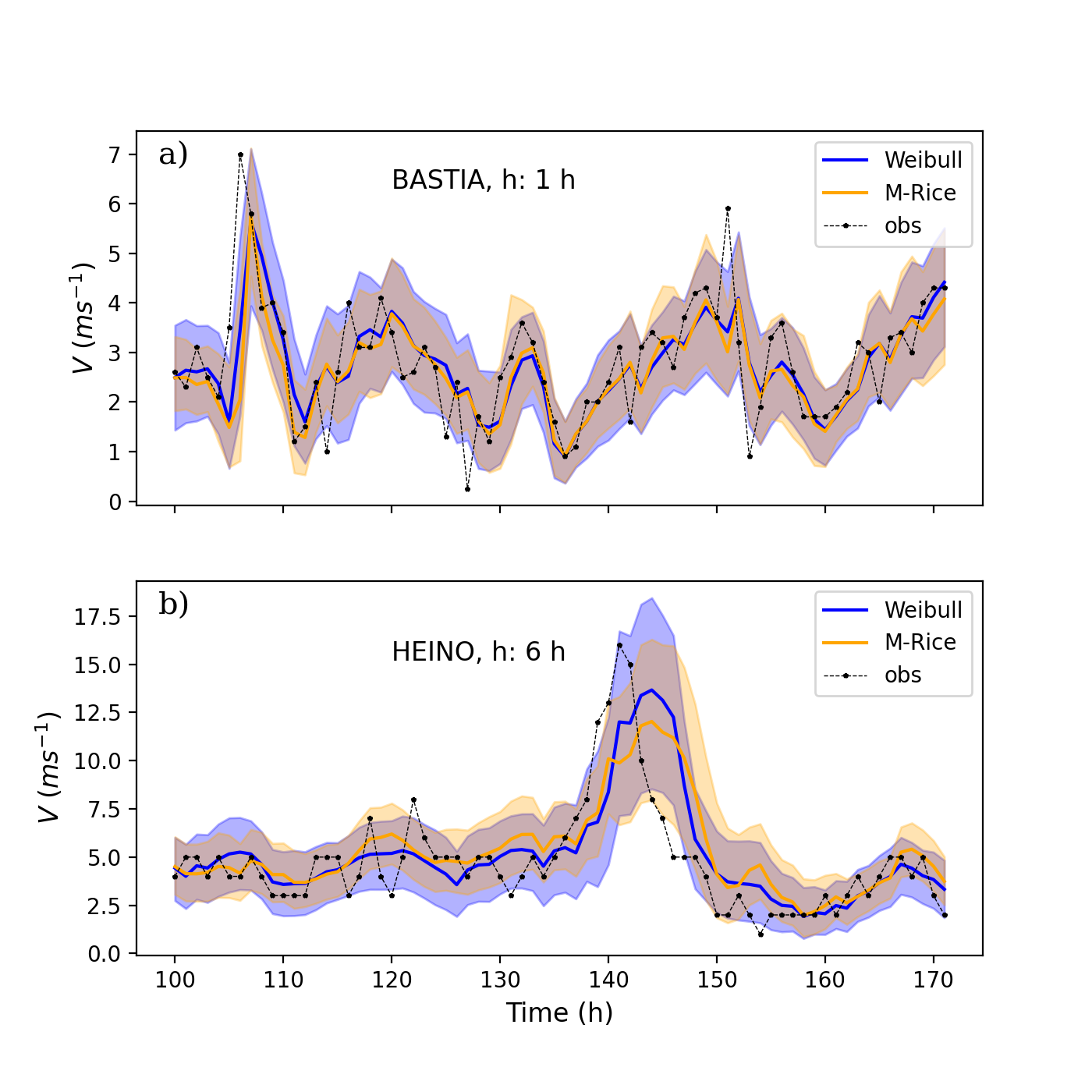}%
	\caption{Illustration of the probabilistic prediction in a) Bastia at $1 h$ horizon and b) Heino at $6 h$ horizon for Weilbull (blue) and M-Rice (orange) distributions.  Shaded regions represent the 80 \% confidence intervals around the median displayed as solid lines.}
	\label{fig_Bastia_Heino}
\end{figure}

In this section, we present the main outcomes of the probabilistic wind speed forecasting method outlined in  section \ref{sec:reg}. As emphasized previously, our primary objective is not to propose the absolute best-performing model specifically tailored for the considered prediction tasks, locations and time-frames. Instead, we aim to unveil the essential features about the differences between the various parametric distributions reviewed in sec. \ref{sec:dist},  the metrics one uses to measure the probabilistic forecasting performances reviewed in sec. \ref{sec:metrics},  and to stress the importance, in the probabilistic prediction accuracy,  of using spatio-temporal information as provided by surrounding stations. Along the same line as in \cite{bamu_23}, to streamline our approach and avoid excessive fine-tuning, we largely adhered to established practices within the literature when selecting model hyperparameters. All the deviations we considered in hyper-parameter values were confined to a limited set of alternatives to these conventional values. We observed that such variations had only marginal impacts on the overall performance metrics discussed below. For all forecasting horizons $h$, the number of considered past time-steps was $n = 3h+1$ meaning that the $4$ previous values of the velocity components $(u,v)$ were considered when $h=1$ hour while $19$ past values were  provided as input when $h=6$ hours.
In all cases a droupout is applied to the inputs of each LSTM layer with a rate $p = 15 \%$ when $h=6$ and $p=2 \%$ when $h=1$. The mini-batch size was chosen to be $B = 512$ in all cases.

\subsection{Probability density function comparison}   
\label{sec:pdf_comparison}
\begin{figure}[H]%
	\centering%
	\includegraphics[width=0.8\linewidth]{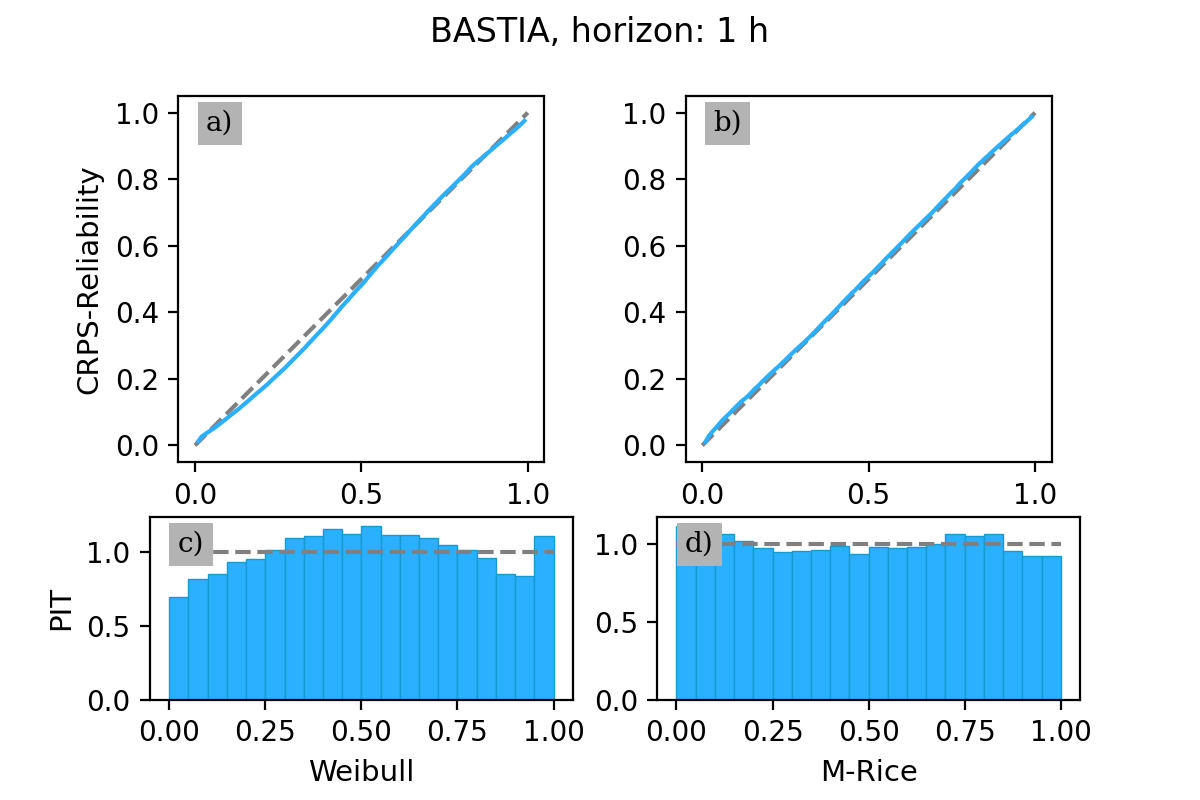}%
	\caption{CRPS-based reliability plots (top panel) and PIT (bottom panel) for Bastia at $h=1$ hour, for the Weibull (left) an M-Rice (right) distributions.}
	\label{fig_Bastia_1}
\end{figure}

\begin{figure}[H]%
	\centering%
	\includegraphics[width=0.8\linewidth]{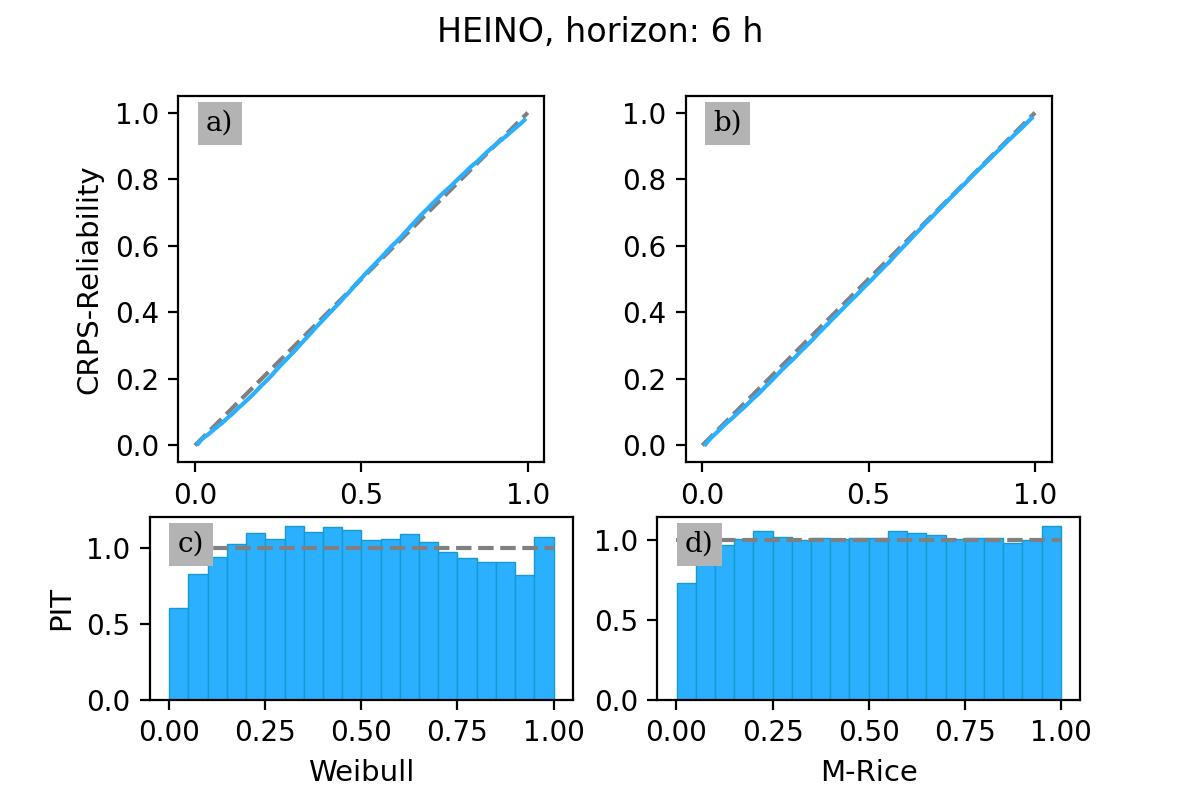}%
	\caption{CRPS-based reliability plots (top panel) and PIT (bottom panel) for Heino at $h=6$ hours, for the Weibull (left) ans M-Rice (right) distributions.}
	\label{fig_Heino_1}
\end{figure}

Our initial objective is to assess the probabilistic predictions stemming from the seven distinct probability density functions outlined in section 3.2. For that purpose, we carried out the first forecasting experiments for sites in Corsica and in the Netherlands using a common model configuration.  The input vector (the ``features'') was built from the past velocities values of the site under consideration and of the 15 closest surrounding sites and, as mentioned previously,  we use the $3h+1$ past values in order to feed the LSTM layers. We focused on 2 different prediction horizons $h$, namely $h=1$ hour and $h=6$ hours. A sample of the results we obtained is provided in Figs \ref{fig_Bastia_Heino},\ref{fig_Bastia_1}  and \ref{fig_Heino_1} where the comparative prediction performances of M-Rice and Weibull distributions are illustrated for Heino and Bastia stations. In Fig. \ref{fig_Bastia_Heino}, we display,  over a time interval of length 72 hours, the observed wind velocity together with a 80\% confidence interval around the median as provided by the M-Rice and Weibull predictions at horizon $h=1$ hour for the Bastia station (top panel) and $h=6$ hours for the Heino station (bottom panel). We see that both methods yield results close to each other with prediction interval values that encompass the observed values very well, except for a few intense events as one would expect. It also clearly appears that the  M-Rice law provides globally sharper confidence intervals and therefore performs slightly better than Weibull law in terms of sharpness. This is confirmed by the Sharp scores (Eq. \eqref{sharpness}) evaluated on the full validation time sample, which are $\text{Sharp}=0.466,\,0.414$ for Weibull and M-Rice respectively, in the case of Bastia $h=1$ hour, $\text{Sharp}=0.512,\,0.472$ for Heino $h=6$ hours (the same trend is observed for the sharpness measures Res defined in \ref{sec:app_1} by Eq. (\ref{def_Rel})). 

M-Rice also performs better in terms of reliability as illustrated clearly in Figs.  \ref{fig_Bastia_1}  and \ref{fig_Heino_1} where are reported examples of plots  that are typically used to qualitatively assess the model prediction reliability. In the top panels we show to so-called CRPS reliability plot that consists in plotting, for a given set of probabilities, $0 \leq p_i = i \Delta p \leq 1$, $o(p_i)$ as defined in Eq. \eqref{def_o}. A perfect prediction corresponds the  diagonal line, $o(p) = p$ reported as the dashed lines in Figs \ref{fig_Bastia_1} et Figs \ref{fig_Heino_1}. In the bottom panels are displayed the PIT histograms as defined in previous section (Eq. \eqref{pit}). In the two considered examples, namely $h=1$ in Bastia and $h=6$ in Heino, one can clearly see that M-Rice appears to provide more reliable predictions than Weibull. A quantitative support to these observations is given by the CRPS and RI scores. Specifically, for the 1-hour horizon in Bastia, M-Rice yields scores of 0.508 (CRPS) and 0.048 (RI), while the Weibull model achieves a score of 0.516 (CRPS) ad 0.106 (RI). Moreover,  for the 6-hour horizon in Heino,  M-Rice demonstrates scores of 0.543 (CRPS) and 0.032 (RI), whereas the Weibull model scores 0.553 (CRPS) and 0.095 (RI).

The overall results we obtained for all considered probability laws are summarized in Table \ref{table1} for horizon $h=1$ hour and in Table \ref{table2} for $h=6$ hours. All the outcomes have been averaged across the four sites within their respective regions. The mean scores of Corsican stations are presented in the left tables, while those of Dutch stations are displayed in the right tables.  Let us recall that the training task of each model consists in minimizing  a loss function that corresponds to the negative log-Likelihood denoted as LogS, defined in Eq. \eqref{def:logS}. The scores that are reported in each table correspond to the LogS, the CRPS and their analogues that focus on extreme values, namely the twCRPS and the censored likelihood CSL that have been defined in Sec. \ref{sec:metrics} (Eqs \eqref{twcrps} and \eqref{csl}).  Supplementary scores, with a specific emphasis on reliability and sharpness, can be found in \ref{sec:app_2}.

A first striking feature that derived from these tables is that for both regions, for almost all scores and all horizons, the method relying on the M-Rice distribution appears to be the most effective one.  This is also evident in all sharpness/resolutions scores and most of reliability scores reported in  Tables \ref{table1_app2} and \ref{table2_app2}.
Table \ref{table3}, summarizes all distribution performances  by providing their mean score ranking. In the left table are given the average ranks of each law, as obtained by considering all stations at both time horizons, while in the right table, for each scoring rule and each rank, we give the corresponding probability distribution. We see that M-Rice appears in the first rank for all considered scores, its average rank  being the closest to 1 in each case. It is noteworthy that the LogS score, which corresponds to the primary objective minimized in the training phase of each model, has average rank 1. This means that for all horizons and all sites in both Corsica and the Netherlands, the M-Rice provides the best log-likelihood over the validation period. Concerning the other scores, we observe that the mean rank of M-Rice is close or smaller than 2, which indicates that it also turns out to be the best method for these scores. One can also remark from the right table \ref{table3}, that, on a general ground, "Rician" distributions provide better performances with respect to the other ones. This may reflect the fact that models based on a conditional Gaussian nature of $(u,v)$ wind components, can accurately account for the fluctuations of the wind intensity. This observation is consistent with the fact, first observed in Ref. \cite{bamu_23} and confirmed within the framework of this paper, that a predicting model with $(u,v)$ components as input features turns out to provide better results than approaches based on wind amplitudes and direction.

It is noteworthy from Tables \ref{table1} and \ref{table2}, that, for all models at small horizons, the scores are better in the Netherlands than in Corsica. This can result from the less homogeneous and more irregular nature of the terrain, which makes surface wind fluctuations less predictable in Corsica at small time horizons. Moreover, at the small scale, each station has its own wind regime (see Fig. \ref{fig_windrose}), the surrounding stations separated by various mountains and relief, are likely to have less correlations with a given site (see the next section). At larger time horizons, the prediction performances for all models are closer to each other, because  the involved wind regimes are of global nature and less dependent on the terrain details. Let us however notice that this is not necessarily true for strong events, as revealed by the twCRPS values. 

\begin{table}[h!]%
	\resizebox{0.5\textwidth}{!}{ %
		\begin{tabular}{c}%
			Corsica\\%
			\\%
			\begin{tabular}{|l|rr|rr|}
				\hline
				\diagbox[width=7em]{Law}{Score}        &   LogS &   CRPS &    CSL &   twCRPS \\
				\hline
				TNormal   &     1.2145 & 0.5060 & 0.1494 & 0.0458 \\
				Weibull   &     1.2152 & 0.5053 & 0.1518 &   0.0449 \\
				Lognormal &     1.2979 & 0.5176 & 0.1517 &   0.0468 \\
				Gamma     &     1.2435 & 0.5106 & 0.1496 &   0.0460 \\
				Nakagami  &     1.2123 & 0.5062 & 0.1467 &   \textbf{0.0448} \\
				Rice      &     1.2038 & 0.5057 & 0.1488 &   \textbf{0.0448} \\
				M-Rice  &     \textbf{1.1758} & \textbf{0.5009} & \textbf{0.1430} &   \textbf{0.0448} \\
				Rayleigh-Rice     &     1.1938 & 0.5038 & 0.1490 &   0.0449 \\
				\hline
			\end{tabular} 
			
	\end{tabular} }  %
	\resizebox{0.5\textwidth}{!}{%
		\begin{tabular}{c}%
			The Netherlands\\%
			\\%
			\begin{tabular}{|l|rr|rr|}
				\hline
				\diagbox[width=7em]{Law}{Score}       &   LogS &   CRPS &    CSL &   twCRPS \\
				\hline
				TNormal   &     1.0828 & 0.4170 & 0.1206 & 0.0274 \\
				Weibull   &     1.0946 & 0.4189 & 0.1249 &   0.0277 \\
				Lognormal &     1.1216 & 0.4206 & 0.1211 &   0.0275 \\
				Gamma     &     1.0928 & 0.4190 & 0.1228 &   0.0278 \\
				Nakagami  &     1.0744 & 0.4175 & 0.1212 &   0.0276 \\
				Rice      &     1.0769 & 0.4173 & 0.1208 &   0.0274 \\
				M-Rice  &     \textbf{1.0685} & \textbf{0.4168} & \textbf{0.1196} &   0.0240 \\
				Rayleigh-Rice     &     1.0747 & 0.4172 & 0.1201 &   \textbf{0.0239} \\
				\hline
			\end{tabular}\\%
	\end{tabular}}%
	\caption{Score values averaged over the four considered sites in Corsica (left) and the four considered sites in the Netherlands (right), for $h=1$ hour and $N=15$.  }%
	\label{table1}
\end{table}

\begin{table}[h!]%
	\resizebox{0.5\textwidth}{!}{%
		\begin{tabular}{c}%
			Corsica\\%
			\\%
			\begin{tabular}{|l|rr|rr|}
				\hline
				\diagbox[width=7em]{Law}{Score}        &   LogS &   CRPS &    CSL &   twCRPS \\
				\hline
				TNormal   &     1.3853 & 0.6090 & 0.1729 & 0.0544 \\
				Weibull   &     1.4365 & 0.6145 & 0.1812 &   0.0534 \\
				Lognormal &     1.5283 & 0.6548 & 0.1922 &   0.0613 \\
				Gamma     &     1.4476 & 0.6135 & 0.1775 &   0.0557 \\
				Nakagami  &     1.4088 & 0.6027 & 0.1729 &   0.0529 \\
				Rice      &     1.4047 & 0.6020 & 0.1759 &   0.0527 \\
				M-Rice  &     \textbf{1.3617} & \textbf{0.5916} & \textbf{0.1680} &   \textbf{0.0522} \\
				Rayleigh-Rice     &     1.4013 & 0.5992 & 0.1772 &   0.0531 \\
				\hline
			\end{tabular}\\%
	\end{tabular}}%
	\resizebox{0.5\textwidth}{!}{%
		\begin{tabular}{c}%
			The Netherlands\\%
			\\%
			\begin{tabular}{|l|rr|rr|}
				\hline
				\diagbox[width=7em]{Law}{Score}       &   LogS &   CRPS &    CSL &   twCRPS \\
				\hline
				TNormal   &     1.4597 & 0.6195 & 0.1827 & 0.0457 \\
				Weibull   &     1.4547 & 0.6160 & 0.1823 &   0.0392 \\
				Lognormal &     1.5011 & 0.6276 & 0.1900 &   0.0410 \\
				Gamma     &     1.4617 & 0.6185 & 0.1856 &   0.0406 \\
				Nakagami  &     1.4388 & 0.6107 & 0.1811 &   0.0396 \\
				Rice      &     1.4404 & \textbf{0.6097} & \textbf{0.1791} &   \textbf{0.0389} \\
				M-Rice  &     \textbf{1.4341} & 0.6099 & 0.1802 &   0.0396 \\
				Rayleigh-Rice     &     1.4435 & 0.6119 & \textbf{0.1791} &   0.0390 \\
				\hline
			\end{tabular}\\%
	\end{tabular}}%
	\caption{Score values averaged over the four considered sites in Corsica (left) and the four considered sites in the Netherlands (right), for $h=6$ hours and $N=15$.  }%
	\label{table2}
\end{table}

\begin{table}[h!]%
 \resizebox{0.475\textwidth}{!}{%
	\begin{tabular}{c}%
		\\%
		\begin{tabular}{|l|cccccccc|r}
			\hline
			\diagbox[width=6 em]{Score}{Rank}     &   TNorm &   Weib &   Logn &   Gamma &   Naka &   Rice &   M-Rice &   RRice \\
			\hline
			LogS  &       4.6 &       5.6 &         8.0 &     6.6 &        3.8 &    3.6 &       {\bf 1.0} &     2.9 \\
			CRPS      &       4.6 &       5.1 &         7.9 &     6.4 &        3.8 &    3.1 &        {\bf 1.9} &     3.2 \\
			RI        &       3.9 &	       5.1&      	7.9	&      6.6	&       4.3&	 2.7&      {\bf 2.4	}&   3.1  \\
			Sharp. &       4.7 &       6.5 &         6.1 &     4.1 &        3.8 &    2.4 &        {\bf 1.4} &     5.1 \\
			\hline
		\end{tabular}%
	\end{tabular}}%
 \resizebox{0.5\textwidth}{!}{
	\begin{tabular}{c}%
		\\%
		\begin{tabular}{|l|cccccccc|l}
			\hline
			\diagbox[width=6 em]{Score}{Rank}     & 0        & 1       & 2        & 3        & 4        & 5       & 6         & 7         \\
			\hline
			LogS  & M-Rice & RRice   & Rice     & Naka & TNorm & Weib & Gamm     & Logn \\
			CRPS      & M-Rice & Rice    & RRice    & Naka & TNorm  & Weib & Gamm     & Logn \\
			RI         & M-Rice	&  Rice	&RRice	&Tnorm	&Naka	&Weibl	&Gamma 	&Logn \\
			Sharp. & M-Rice & Rice    & Naka & Gamm    & TNorm  & RRice   & Logn & Weib   \\
			\hline
		\end{tabular}%
	\end{tabular}}%
	\caption{Ranking of all method performances calculated on all the considered sites and horizons (N = 15)}%
	\label{table3}
\end{table}

\subsection{Improvement obtained using data from surrounding locations}
\begin{figure}[tbh]
	\centering
	\hspace*{0.0cm}
	\includegraphics[width=0.95\columnwidth]{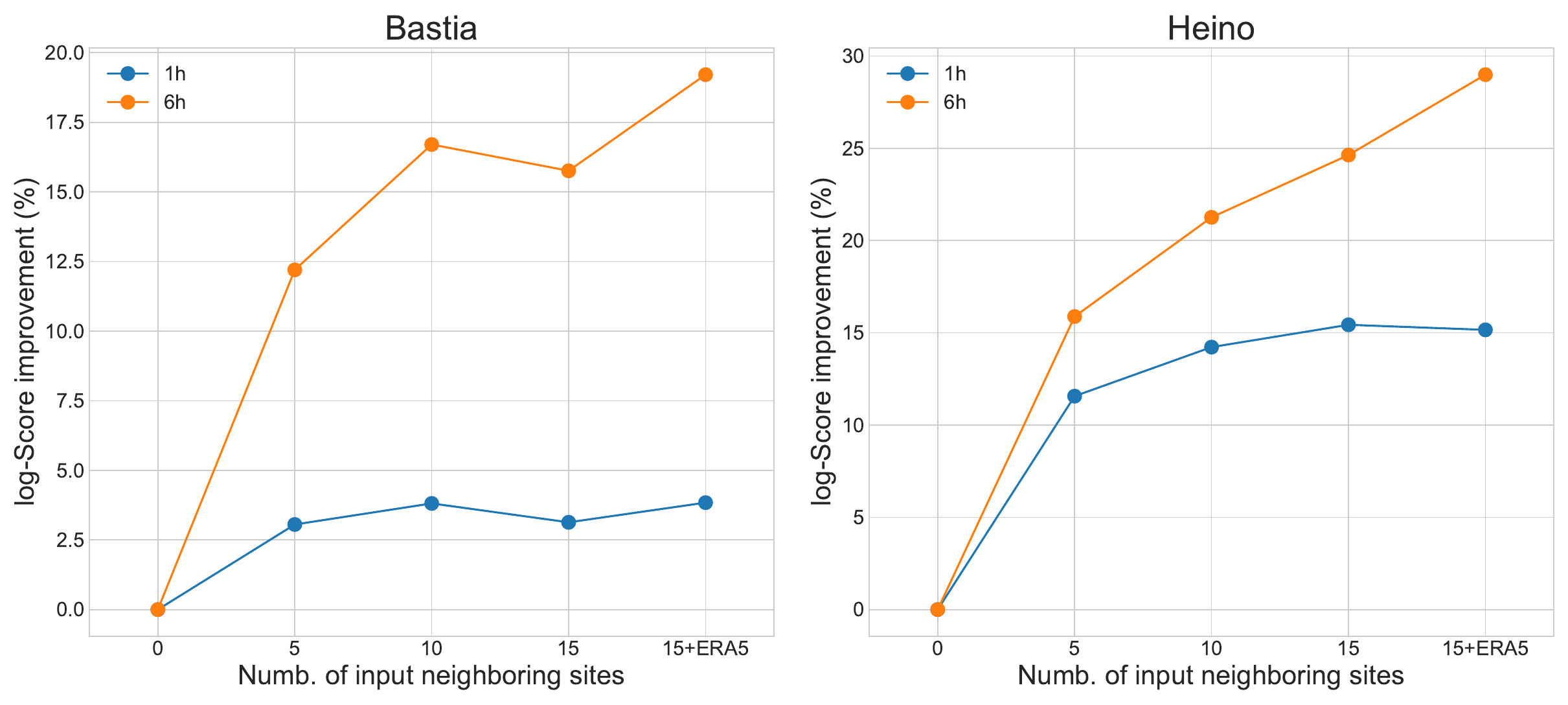}
	\caption{M-Rice LogS improvement as a function of the considered input information at Bastia (left panel) and Heino (right panel). As the number of input sites increases, one 
		observes better prediction performances with a quick saturation at the shortest prediction horizon.}
	\label{fig_logscore}
\end{figure}

\begin{figure}[tbh]
	\centering
	\hspace*{0.0cm}
	\includegraphics[width=0.95\columnwidth]{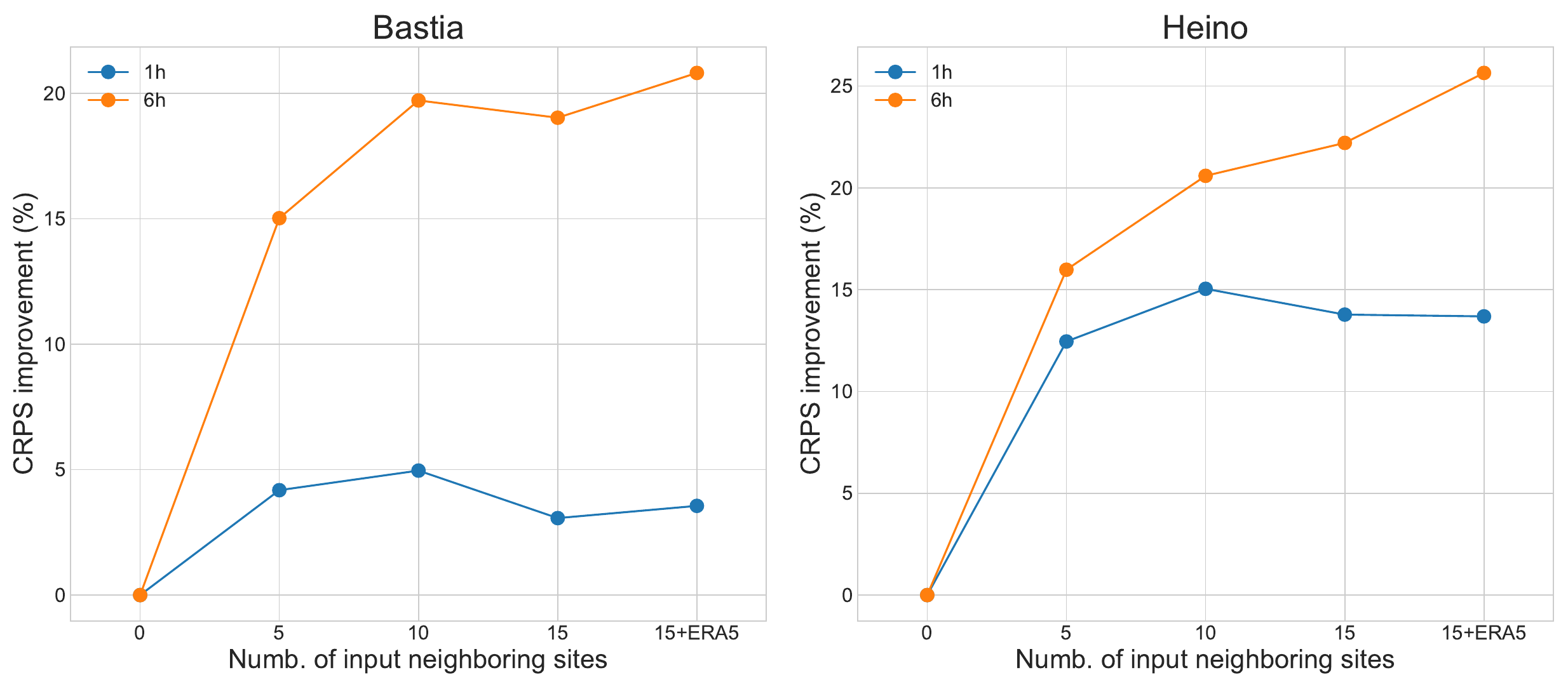}
	\caption{M-Rice CRPS improvement as a function of the considered input information at Bastia (left panel) and Heino (right panel). As the number of input sites increases, one 
		observes better prediction performances with a quick saturation at the shortest prediction horizon.}
	\label{fig_crps}
\end{figure}
Beyond the results of the previous section showing that the methods based on Rice law, especially the M-Rice model, are the most effective, we remark that most approaches have scores that are very close to each other. This is particularly true at 1 hour horizon, for CRPS,  twCRPS and CSL scores (see Table \ref{table1}). This observation can be explained by the fact that the Rice distribution, within a range of parameters corresponding to the ones observed empirically, can be closely approximated by other pdf classes like Gamma, Weibull or Nakagami (for example, in \cite{marqum_approx} an accurate approximation of the Rice cumulative distribution is built using a Weibull shape).
In this section,  we show that accounting for the information related to the spatial as well as temporal distribution of wind speed, can represent a  more significant improvement than the choice of the "correct" shape of the underlying probability distribution.

Let us start by illustrating the benefits of using, as input to a given model,  wind speed data from an increasing number of nearby stations, for 2 reference sites, namely Bastia and Heino. In the case of M-Rice probabilistic forecasting, we estimate the  improvement of the performances as measured by the LogS and the CRPS scores, when one is taking into account respectively  the wind speed components of the 5, 10, 15 closest stations and finally by also taking into account the wind speed at the 15 closest ERA5 nodes. The results are reported in Fig. \ref{fig_logscore} for the LogS and Fig. \ref{fig_crps} for the CRPS.
In each case, we plot the improvement (in percent) as respect to the "basic" prediction that only takes into account "local" information, namely the past values of the wind speed at the considered station\footnote{Notice that in the case of CRPS, we report the relative variations while in the case of the log-Score, since it is defined from the logarithm of the likelihood, we just simply report its variations.}. One can check that the curves associated with 
these 2 scoring rules display the same behavior. On can also clearly see that all curves are increasing meaning that, whatever the considered horizon and for both sites, using more and more information from neighboring locations statistically enhances our capacity to predict the distribution of upcoming fluctuations. Notice that similarly to what has been observed for the mean square error in the context of deterministic forecasting in Ref. \cite{bamu_23}, we observe a rapid saturation of performance gains at the 1-hour horizon. One can see that, as the forecast horizon increases, integrating information from remote sites as input into the model becomes increasingly valuable. Indeed, as emphasized in \cite{bamu_23}, locations in the upwind direction are expected to be of utmost importance in predicting future weather conditions, as they are, in a sense, advected by the primary surface wind. In that respect, accounting for the wind speed values at too distant sites is useless at short horizons. Moreover, it can be observed that in Bastia, the contribution of surrounding stations impacts less the overall performances than what observed for the site of Heino in the Netherlands, especially when considering the 1 hour horizon: the plateau occurs faster and the highest improvement of the score is smaller than what observed in Heino. This is due to the peculiarity of Corsica island which has a more complex orography than the 
Netherlands: Corsica is characterized by its rugged, mountainous terrain, with a wider variety of local weather conditions while the Netherlands are predominantly flat and low-lying resulting in a more homogeneous weather situation. As reflected by a greater similarity between the score behavior in Bastia and Heino at horizon 6 hours, the impact of the local orography becomes less important when considering long prediction horizons, that is, terrain details have less impact than large-scale meteorological patterns.
One can remark that at 6 hours horizon, prediction quality continuously increases as one account for more and more
distant information and also significantly benefits from ERA5 reanalysis data at nearby grid points.

\begin{table}[h!]%
	\resizebox{0.5\textwidth}{!} {%
		\begin{tabular}{c}%
			Corsica\\%
			\begin{tabular}{|l|c|c|c|c|}
				\hline
				\diagbox[width=7em]{N}{Score}     &   LogS &   CRPS &    CSL &   twCRPS \\
				\hline
				15      &      3.19 \% &  3.08 \% & 5.74 \% &    4.28 \% \\
				15+ERA5 &      3.75 \% &  3.71 \% & 6.77 \% &    5.46 \% \\
				\hline
			\end{tabular}\\
		\end{tabular}%
	}%
	\resizebox{0.5\textwidth}{!} {%
		\begin{tabular}{c}%
			The Netherlands\\%
			\begin{tabular}{|l|c|c|c|c|}
				\hline
				\diagbox[width=7em]{N}{Score}     &   LogS &   CRPS &   CSL &   twCRPS \\
				\hline
				15      &     15.93 \% & 14.67 \% & 16.45 \% &   16.67 \% \\
				15+ERA5 &     15.83 \% & 14.66 \% & 16.56 \% &   16.53 \% \\
				\hline
			\end{tabular}\\%
		\end{tabular}%
	}%
	\caption{M{-}Rice  relative score improvements at $h=1$ hour when one considers, other than the reference site, additional wind speed information from 15 neighboring stations or from 15 neighboring station together with ERA 5 data at 15 closest grid points.}%
	\label{table4}
\end{table}

\begin{table}[h!]%
	\begin{footnotesize}
		\resizebox{0.5\textwidth}{!} {%
			\begin{tabular}{c}%
				Corsica\\%
				\begin{tabular}{|l|c|c|c|c|}
					\hline
					\diagbox[width=7em]{N}{Score}     &   LogS &   CRPS &    CSL &   twCRPS \\
					\hline
					15      &     18.55 \% & 19.84 \% & 24.95 \% &   29.31 \% \\
					15+ERA5 &     23.00 \% & 22.45 \% & 27.60 \% &   30.61 \% \\
					\hline
				\end{tabular}\\%
			\end{tabular}%
		}%
		\resizebox{0.5\textwidth}{!} {%
			\begin{tabular}{c}%
				The Netherlands\\%
				\begin{tabular}{|l|c|c|c|c|}
					\hline
					\diagbox[width=7em]{N}{Score}     &   LogS &   CRPS &    CSL &   twCRPS \\
					\hline
					15      &     26.05 \% & 23.13 \% & 19.97 \% &   18.87 \% \\
					15+ERA5 &     31.53 \% & 27.36 \% & 23.86 \% &   22.56 \% \\
					\hline
				\end{tabular}\\%
			\end{tabular}%
		}%
		\caption{M{-}Rice relative score improvements at horizon 6 hours when one considers s, other than the reference site, additional wind speed information from 15 neighboring stations or from 15 neighboring station together with ERA 5 data at 15 closest grid points.}%
		\label{table5}
	\end{footnotesize}
\end{table}

The results observed for Bastia and Heino are confirmed at other locations. Tables \ref{table4} and \ref{table5}
summarize our findings in the case of M-Rice distribution. For both regions and time horizons, we report the mean relative score improvement, as respect to a model that only utilizes local wind velocities and seasonal features $\bS_t$,  when one accounts for the past wind velocities at the 15 closest stations and when one also accounts for the ERA5 data (as described in Sec \ref{sec:data}) at the 15 grid points surrounding the considered station. As shown in \ref{sec:app_3} for Weibull law, similar results can be obtained when one uses alternative distributions. 
The relative performances reported in these tables first confirm that accounting for  spatio-temporal information always increases the forecasting score. It is however worth noting that ERA5 data do not allow for any improvement in scores for $h=1$ hour. This is not surprising, as reanalysis data, constructed using numerical simulations on a relatively coarse grid, do not capture fine-scale orographic details and local phenomena. Therefore, they are not expected to reveal new dynamic features compared to data recorded at nearby stations. Such an observation results from the same ``plateau effect" as observed in  Fig. \ref{fig_logscore} and Fig. \ref{fig_crps} at horizon $h=1$ hour. A indicated by the results in table \ref{table4}, the situation is different at larger horizons: when $h=6$ hours,  one can see that reanalysis data allow one to really improve all scores.

The results reported in Table \ref{table4} also confirm something we observed and commented previously: Due to the very different orographic and geographical conditions, the increase in performance in Corsica by incorporating the situation of neighboring sites is potentially much lower than in the Netherlands for short-term horizons. As it can be checked in Table \ref{table5}, the differences between the two regions become less pronounced at larger time horizons where, as emphasized previously, the large-scale weather conditions play a much more significant role.

From Tables \ref{table1_app2} and \ref{table2_app2} of \ref{sec:app_2}, one can remark that CRPS reliability scores of all methods are very small as compared to the resolution scores. This means that the reliability of the considered laws are  "very good"
as compared to their resolution that dominate in the final CRPS value. Along the same vein, one might wonder 
whether the improvement in CRPS score is primarily due to the enhancement of reliability or if it is associated with the gain in resolution when adding information to regress the parameters of the considered distribution. Table \ref{table6} answer this question for the M-Rice distribution. The values of the contribution of the Reliability and the Resolution terms in the CRPS decomposition  (see \ref{sec:app_1})  are reported for both regions and horizons. It appears that including more information in the prediction model does not improve (or improves only marginally) the reliability while the resolution is significantly better. This observation is rather intuitive since one can think that the M-Rice estimation tries to approximate the  wind speed real conditional law in each situation. This latter depends on the considered information only through its resolution while its reliability is always perfect. 

Let us finally observe from Tables \ref{table4} and \ref{table5} that 
the improvements in the scores of the extremes, namely CSL and twCPRS, are of the same order of magnitude as their non-conditional counterparts. Issues specifically related to probabilistic forecasting of extreme events deserve a specific study that is beyond the scope of the present paper and will be the object of a future work.

\begin{table}[h!]%
\centering
		\begin{tabular}{c}%
			Corsica\\%
			\begin{tabular}{|l|c|c|c|c|}
				\hline
				\diagbox[width=7em]{Horizon}{Score}     &   Rel &   Res \\
				\hline
				
				1 h &     0.0 \% &      3.7 \% \\
				6 h &      0.5 \% &     22 \%  \\
				\hline
			\end{tabular}\\%
			\newline%
		\end{tabular}%
		\begin{tabular}{c}%
			The Netherlands\\%
			\begin{tabular}{|l|c|c|c|c|}
				\hline
				\diagbox[width=7em]{Horizon}{Score}     &   Rel  &   Res  \\
				\hline
				1 h &      0.0 \% &     14.7\% \\
				6 h &      2.0 \% &     25.4 \% \\
				\hline
			\end{tabular}\\%
			\newline%
		\end{tabular}%
	\caption{Respective contribution of the increase of Reliability and Resolution to the M{-}Rice  relative CRPS improvements when one adds $N = 15$ neighbor sites and ERA5 wind data information. $\Delta Rel/CRPS_{N=0}$ and  $\Delta Res/CRPS_{N=0}$ are reported for both regions and time horizons.}
	\label{table6}%
\end{table}

\subsection{Point forecast performances}
Along the main lines presented in Sec. \ref{sec:pforecast}, one can also estimate the performances of the probabilistic predictions in their capacity to provide ``point", i.e., deterministic, forecasts.
Given that we have already performed, in the previous sections, a comparison of different distributions using a comprehensive set of suitable metrics, in the results presented below, we only consider the M-Rice distribution, which exhibited the best performance. We compare the ability of the latter to produce the best point forecast in terms of MAE (median value) or RMSE (mean value) against two classical reference models described in Sec. \ref{sec:pforecast}, namely persistence, and linear models. The comparison has been conducted using the information at the considered station and all its $N=15$ neighbor sites.
The linear regression, implemented using the “LinearRegression” model from the Python
Scikit-Learn library \cite{Pedregosa:2011}, uses as input the same set of past wind speed values as the Neural network model. The results are reported in Table \ref{table7} below. One can see that the probabilistic approach gives for the two regions and both considered time horizons, significantly better results than the two reference models. We have noticed that for all sites in the Netherlands, the probabilistic approaches based on the LSTM model, provide 
point forecasts with comparable accuracy as the best Neural Network models specifically
trained for that purpose in \cite{bamu_23}.   

\begin{table}[H]%
	\centering
		\begin{tabular}{c}%
			Corsica\\%
			\begin{tabular}{|l|c|c|c|c|}
				\hline
				\diagbox[width=6em]{Model}{horizon}     &  1h &  6h \\
				\hline
				
				Persistence  &   0.86 / 1.22  &     1.68 / 2.27 \\
				Linear &    0.79 /1.10 &     1.31 / 1.76  \\
				M-Rice  &    {\bf 0.70} / {\bf 0.99}  &      {\bf 0.81} / {\bf 1.16}  \\
				\hline
			\end{tabular}\\%
			\newline%
		\end{tabular}%
			\begin{tabular}{c}%
			The Netherlands \\%
			\begin{tabular}{|l|c|c|c|c|}
				\hline
				\diagbox[width=6em]{Model}{horizon}     &  1h &  6h \\
				\hline
				Persistence  &   0.67 /1.01  &    1.42 / 1.90 \\
				Linear &    0.62 / 0.82 &     1.17 / 1.50  \\
				M-Rice  &    {\bf 0.58} / {\bf 0.70}  &      \bf {0.84} / {\bf 1.14}  \\
				\hline
			\end{tabular}\\%
			\newline%
		\end{tabular}%
	\caption{MAE et RMSE performances in point forecast of the M-Rice distribution compared to Persistence and linear models. The values of the average MAE / RMSE  in $ms^{-1}$ for the four tested stations in Corsica (left table) and The Netherlands (right table) are reported for both prediction horizons $h=1$ hour and  $h=6$ hours.}
	\label{table7}%
\end{table}

\section{Conclusion}
\label{s_conc}

This study delves into the realm of intelligent weather forecasting by exploring the application of parametric probabilistic forecasting techniques to wind speed prediction. Leveraging hourly wind data from monitoring stations in the Netherlands and Corsica (France), we compare the performances of various well known distributions for wind speed fluctuations including Truncated Normal,  Weibull, Gamma and two recently introduced  models based on the Rice distribution. We notably consider the so-called "Multifractal-Rice" (M-Rice) distribution introduced in \cite{BaMuPo10c} within the context of a stochastic model of wind fluctuations inspired from random cascade models of fully developed turbulence. It turns out that this distribution consistently demonstrates superior performance across different forecasting horizons, weather stations, and evaluation metrics.

Our investigations also encompass the influence of spatial information integration in a probabilistic forecasting framework.
 To this end, we assess the collaborative impact of neighboring station data, as well as the inclusion of supplementary information derived from wind speed values provided by reanalysis data (ERA5) at adjacent grid points.
We furnish compelling evidence in favor of leveraging wind speed data from neighboring stations.
Similar to point prediction, our findings demonstrate a consistent improvement in probabilistic forecasting accuracy with the addition of data from nearby sites. By decomposing the CRPS score in Reliability and Resolution contributions, we found that the increase in probabilistic forecasting performances is primarily due to the increase in resolution, the reliability being very good in all cases. Furthermore, we note that as the forecasting horizon lengthens, the incorporation of data from geographically distant locations or the consideration of reanalysis ERA5 data becomes increasingly beneficial.
The enhancement in probabilistic metrics such as LogS or CRPS values can reach levels approaching 30\% for large forecasting horizons ($h=6$ hours). For shorter horizons (h=1 hour), we observed that the accuracy of probabilistic forecasting, although improved, is relatively lower, especially in Corsica, where short-term meteorological changes are harder to predict than in the Netherlands due to its more complex geographical and orographic conditions.

In a future approach, we intend to utilize, in addition to surrounding weather sensor data,  the whole spatio-temporal fields of wind speed (and other atmospheric variables) as provided by the output of the high-resolution numerical weather prediction model in order to further improve the forecasting performances at short-term horizons. 

Finally, as discussed in Sec. \ref{sec:data}, an important issue, critical for energy production purpose, concerns  the reliability of the approaches advocated in this paper for predicting wind speed, and thereby wind power, at various heights, notably at heights 100m and beyond.  
We believe that the present approach, that sheds light on the interplay between spatial data collaboration and probabilistic distribution selection, offers insights at the surface level that are likely to provide efficient methodologies applicable at the elevated altitudes involved in wind energy production. These aspects will be addressed in a forthcoming study by leveraging specific wind data record at hub-heights. Thereby, we also hope to confirm the pertinence of commonly used conversion rules and assert the broad applicability of our work that leverages surface-level data for the development of robust wind energy forecasting tools.

\section*{Acknowledgment}
This work was partially supported by ANR grant SAPHIR project ANR-21-CE04-0014-03.

\appendix

\section{Hersbach decomposition of the CRPS}
\label{sec:app_1}
In this section, we recall how, as first shown by Hersbach \cite{hersbach2000decomposition}, one can decompose the CRPS defined in Eq. \eqref{crps} 
as a sum of 3 terms representing respectively the Reliability, the Resolution and the Uncertainty. 
In \cite{hersbach2000decomposition}, the author consider a discrete estimation of the CRPS at each time step $k$  relying the observation of $N$ ensemble members $x^{k}_1, \ldots, x^{k}_N$ for each observation time
$k = 1, \ldots, M$.   Hereafter, along the same line as the approach developed in the Appendix of \cite{candille2005evaluation}, we provide a continuous analogue of Hersbach decomposition that is adapted 
to the framework of this paper since it involves the cumulated probability distribution $F_k(z)$ at each time instead of an ordered set of $N$ draws with this law.

Let us denote by $\{O_k \}_{k=1 \ldots M}$ the different ``states of nature'' observed at times $k=1,2,\ldots,M$. For any  function  $f(O)$ that depends on the state $O$, by denoting $f_k = f(O_k)$, we can define the mean value over the observations as:
$$
\E(f_k) = \frac{1}{M} \sum_{k=1}^M f_k \; .
$$

Let $F_k(z) = P_k(v \leq z)$ be the predicted wind speed cumulated distribution function  and $y_k$ be the observed values  at time $k$. The CRPS at time $k$ is defined as:
$$
CRPS_k = \int_0^{+\infty} \Big( F_k(z)-H(z-y_k) \Big)^2 dz
$$
where $H(x)$ stands for the Heaviside function.
If we assume that $F_k(z)$ is continuous and strictly increasing on $[0,\infty)$, 
one has:
$$
CRPS_k = \int_0^{+\infty} \Big[ F_k(z)-H \Big( F_k(z)-F_k(y_k) \Big) \Big]^2 dz
$$
and following \cite{candille2005evaluation}, one can perform the change of variable $p =  F_k(z)$ (i.e. $z =F_k^{-1}(p)$),
to obtain:
\begin{equation}
	\label{crps1}
	CRPS_k = \int_0^1 \Big(p-H(p-p_k) \Big)^2 \gamma_k(p) dp
\end{equation}
with
$\gamma_k(p) = \frac{d}{dp}F^{-1}_{k}(p)$, $p_k = F_k(y_k)$.
Let define $\alpha_k(p)$ and $\beta_k(p)$ as follows: 
\begin{equation}
	\label{def_alpha_beta}
	\beta_k(p)  =  \gamma_k(p) H(p-p_k) \; \; , \; \;
	\alpha_k(p)   =  \gamma_k(p)\Big(1-H(p-p_k)\Big) \; .
\end{equation}
Note that we have obviously:
$$\alpha_k(p)+\beta_k(p) = \gamma_k(p) \;,$$
and Eq. \eqref{crps1} can be rewritten as:
\begin{equation}
	\label{crps2}
	CRPS_k = \int_0^1 \Big( \alpha_k(p) p^2 + (1-p)^2 \beta_k(p) \Big) dp
\end{equation}
which is the continuous analog of Eqs. (24,25) of \cite{hersbach2000decomposition}.
Let us finally define
\begin{equation}
	\label{def_gamma}
	\gamma(p) = \E \Big[\gamma_k(p) \Big] \; , \; \; \alpha(p) = \E \Big[\alpha_k(p) \Big] \;,  \; \; \beta(p) = \E \Big[\beta_k(p) \Big] \; 
\end{equation}
and
\begin{equation}
	\label{def_o}
	o(p) = \frac{\E \Big[ \beta_k(p) \Big]}{\gamma(p)} \; .
\end{equation}
Since 
$$
\alpha(p) = (1-o(p)) \gamma(p)  \; \mbox{and} \;   \beta(p) = o(p) \gamma(p)   \; ,
$$
from \eqref{crps2},
$$CRPS = \E \Big(CRPS_k \Big) $$ can be written as:
\begin{eqnarray}
	\nonumber
	CRPS & = & \int_0^1 \gamma(p) \Big( (1-o(p)) p^2 + (1-p)^2 o(p)  \Big) dp \\
	\label{decomp1}
	& = &  \int_0^1 \gamma(p) \Big(p-o(p) \Big)^2 dp + \int_0^1 \gamma(p) o(p) \Big( 1-o(p) \Big) dp 
\end{eqnarray}

Let us consider that special case when $F_k(z)$ does not depend on $k$ and reads $F_k(z) = F_c(z)$, 
where $F_c(z)$ corresponds to the "climatological" cumulative distribution function defined as 
$$F_c(z) = \E (H(z-y_k)) . $$
In that situation, if we denote by $CRPS_c$ the corresponding CRPS, a direct computation shows that:
$$
CRPS_c = \int_0^{+\infty} F_c(z) (1-F_c(z)) dz = \int_0^1 p(1-p) \gamma_c(p) dp = \int_0^1 o_c(p)(1-o_c(p))  \gamma_c(p) dp
$$
with $o_c(p) = p$ and $\gamma_c(p) = \frac{d}{dp}F^{-1}_{c}(p)$. This means that in this case,
the decomposition \eqref{decomp1} reduces to its last term.
If one defines,  according to Hersbach, the ``uncertainty'' as such ``climatological'' CRPS, 
\begin{equation}
	\label{def_U}
	U = CRPS_c
\end{equation}
the ``resolution'',
\begin{equation}
	\label{def_Res}
	Res = U - \int_0^1 o(p)(1-o(p))  \gamma(p) dp
\end{equation}
and the ``reliability'' term as:
\begin{equation}
	\label{def_Rel}
	Rel =   \int_0^1  \Big(p-o(p) \Big)^2 \gamma(p) dp 
\end{equation}
then Eq. \eqref{decomp1} leads to the Hersbach decomposition:
$$
CRPS = Rel - Res + U \; .
$$

\section{Resolution and sharpness scores}
\label{sec:app_2}
\begin{table}[h!]%
	\resizebox{0.5\textwidth}{!}{
		\begin{tabular}{c}%
			Corsica\\%
			\\%
			\begin{footnotesize}%
				\begin{tabular}{|l|rr|rr|}
					\hline
					Law       &   Rel  &     RI &   Res  &   Sharp \\
					\hline
					TNormal   & 7.5870$\; 10^{-4}$ & 0.0733 &     0.5145 &      0.4592 \\
					Weibull   & 8.9848$\; 10^{-4}$ &  0.1000  &     0.5152 &      0.4577 \\
					Lognormal & 7.4155$\; 10^{-3}$ & 0.2182 &     0.5094 &      0.4612 \\
					Gamma     & 2.7147$\; 10^{-3}$ & 0.1472 &     0.5118 &      0.4508 \\
					Nakagami  & 1.3295$\; 10^{-3}$ & 0.1008 &     0.5148 &      0.4526 \\
					Rice      & 8.0735$\; 10^{-4}$ &  0.0533  &     0.5147 &      0.4360 \\
					M-Rice  & \textbf{4.0070$\; 10^{-4}$} &\textbf{0.0496} &     \textbf{0.5192} &      \textbf{0.4190} \\
					Rayleigh-Rice     & 8.5183$\; 10^{-4}$ &0.0524 &     0.5167 &      0.4470 \\
					\hline
				\end{tabular}%
			\end{footnotesize}%
		\end{tabular}%
	}	
	\resizebox{0.5\textwidth}{!}{
		\begin{tabular}{c}%
			The Netherlands\\%
			\\%
			\begin{footnotesize}%
				\begin{tabular}{|l|rr|rr|}
					\hline
					Law       &   Rel  &     RI &   Res  &   Sharp \\
					\hline
					TNormal   & \textbf{4.7086$\; 10^{-4}$} & 0.0593&     0.7583 &      \textbf{0.3811} \\
					Weibull   & 1.7523$\; 10^{-3}$ & 0.1308&     0.7576 &      0.4047 \\
					Lognormal & 2.3552$\; 10^{-3}$ & 0.1531 &     0.7565 &      0.3915 \\
					Gamma     & 1.4254$\; 10^{-3}$ & 0.1149 &     0.7572 &      0.3899 \\
					Nakagami  & 9.5379$\; 10^{-4}$ & 0.0818&     0.7582 &      0.3848 \\
					Rice      & 6.3992$\; 10^{-4}$ &0.0759 &     0.7581 &      0.3827 \\
					M-Rice  & 5.4932$\; 10^{-4}$ & \textbf{0.0467} &     \textbf{0.7585} &       0.3826 \\
					Rayleigh-Rice     & 9.8896$\; 10^{-4}$ & 0.0695 &     \textbf{0.7585} &      0.3952 \\
					\hline
				\end{tabular}%
			\end{footnotesize}%
	\end{tabular}}
	\caption{Reliability and sharpness scores values averaged over the four considered sites in Corsica (left) and the four considered sites in the Netherlands (right), for $h=1$ hour and $N=15$. }
	\label{table1_app2}%
\end{table}

\begin{table}[h!]%
	\resizebox{0.5\textwidth}{!}{
		\begin{tabular}{c}%
			Corsica\\%
			\\%
			\begin{footnotesize}%
				\begin{tabular}{|l|rr|rr|}
					\hline
					Law       &   Rel &     RI &   Res &   Sharp \\
					\hline
					TNormal   & 1.3449$\; 10^{-3}$ & 0.0684 &     0.4147 &      0.5550 \\
					Weibull   & \textbf{1.1203$\; 10^{-3}$} &0.0631 &     0.4089 &      0.5277 \\
					Lognormal & 6.8048$\; 10^{-3}$ & 0.1849  &     0.3743 &      0.5393 \\
					Gamma     & 2.4907$\; 10^{-3}$ &0.1130  &     0.4113 &      0.5032 \\
					Nakagami  & 1.1622$\; 10^{-3}$ &  0.0557 &     0.4207 &      0.5080 \\
					Rice      & 1.8942$\; 10^{-3}$ & \textbf{0.0499} &     0.4221 &      0.4909 \\
					M-Rice  & 2.2528$\; 10^{-3}$ &0.0824 &     \textbf{0.4329} &      \textbf{0.4390} \\
					Rayleigh-Rice     & 3.0673$\; 10^{-3}$ & 0.0752 &     0.4261 &      0.5128 \\
					\hline

				\end{tabular}%
			\end{footnotesize}%
	\end{tabular}}
	\resizebox{0.5\textwidth}{!}{
		\begin{tabular}{c}
			The Netherlands\\%
			\\%
			\begin{footnotesize}%
				
				\begin{tabular}{|l|rr|rr|}
					\hline
					Law       &   Rel &     RI &   Res &   Sharp \\
					\hline
					TNormal   & 1.2451$\; 10^{-3}$ & 0.0572 &     0.5614 &      0.5589 \\
					Weibull   & 1.1980$\; 10^{-3}$ & 0.0693  &     0.5648 &      0.5738 \\
					Lognormal & 4.3347$\; 10^{-3}$ & 0.1549 &     0.5563 &      0.5641 \\
					Gamma     & 3.1393$\; 10^{-3}$ & 0.1114&     0.5642 &      0.5601 \\
					Nakagami  & 1.6012$\; 10^{-3}$ & 0.0691 &     0.5705 &      0.5587 \\
					Rice      & \textbf{5.3383$\; 10^{-4}$} &0.0503  &     0.5704 &      0.5507 \\
					M-Rice  & 9.3537$\; 10^{-4}$ & \textbf{0.0297} &     \textbf{0.5706} &      \textbf{0.5334} \\
					Rayleigh-Rice     & 8.8015$\; 10^{-4}$ &0.0502&     0.5685 &      0.5657 \\
					\hline
					
				\end{tabular}%
			\end{footnotesize}%
	\end{tabular}}%
	\caption{Reliability and sharpness scores values averaged over the four considered sites in Corsica (left) and the four considered sites in the Netherlands (right), for $h=6$ hours and $N=15$.}%
	\label{table2_app2}
\end{table}

\section{Weibull relative scores}
\label{sec:app_3}
\begin{table}[H]%
\resizebox{0.5\textwidth}{!} {%
\begin{tabular}{c}%
Corsica\\%
\begin{tabular}{|l|c|c|c|c|}
\hline
 \diagbox[width=7em]{N}{Score}     &   LogS &   CRPS &   CSL &   twCRPS \\
\hline
 15      &      3.66 \% &  3.39 \% & 7.48 \% &    4.84 \% \\
 15+ERA5 &      4.24 \% &  3.61 \% & 7.71 \% &    4.61 \% \\
\hline
\end{tabular}\\%
\end{tabular}%
}%
\resizebox{0.5\textwidth}{!} {%
\begin{tabular}{c}%
The  Netherlands\\%
\begin{tabular}{|l|c|c|c|c|}
\hline
 \diagbox[width=7em]{N}{Score}     &   LogS &   CRPS &    CSL &   twCRPS \\
\hline
 15      &     13.34 \% & 14.93 \% & 16.44 \% &   16.94 \% \\
 15+ERA5 &     12.97 \% & 14.73 \% &  9.67 \% &   14.62 \% \\
\hline
\end{tabular}\\%
\end{tabular}%
}%
\caption{Weibull relative score improvements at $h=1$ hour when one considers, other than the reference site, additional wind speed information from 15 neighboring stations or from 15 neighboring station together with ERA 5 data at 15 closest grid points.}%
\end{table}
\begin{table}[H]%
\resizebox{0.5\textwidth}{!} {%
\begin{tabular}{c}%
Corsica\\%
\begin{tabular}{|l|c|c|c|c|}
\hline
 \diagbox[width=7em]{N}{Score}     &   LogS &   CRPS &    CSL &   twCRPS \\
\hline
 15      &      9.53 \% & 17.42 \% & 26.37 \% &   26.98 \% \\
 15+ERA5 &     13.29 \% & 19.73 \% & 33.94 \% &   26.86 \% \\
\hline
\end{tabular}\\%
\end{tabular}%
}%
\resizebox{0.5\textwidth}{!} {%
\begin{tabular}{c}%
The  Netherlands\\%
\begin{tabular}{|l|c|c|c|c|}
\hline
 \diagbox[width=7em]{N}{Score}     &   LogS &   CRPS &    CSL &   twCRPS \\
\hline
 15      &     15.02 \% & 23.09 \% & 23.28 \% &   16.77 \% \\
 15+ERA5 &     17.82 \% & 27.01 \% & 29.73 \% &   20.38 \% \\
\hline
\end{tabular}\\%
\end{tabular}%
}%
\caption{Weibull relative score improvements at $h=6$ hour when one considers, other than the reference site, additional wind speed information from 15 neighboring stations or from 15 neighboring station together with ERA 5 data at 15 closest grid points.}%
\end{table}

\section{Fully connected and linear regime switching models}
\label{sec:app_4}
\subsection{A model composed with fully connected layers}
A simple alternative to the LSTM model described in Section \ref{sec:models}, is a model
that takes as input a single vector $\bX'_{t,N_t}$ that stacks all components of the matrix $\bX_{t,N_{t}}$ defined in Eq. \eqref{def:XtN}:
$$
\bX'_{t,N_t}= \left. 
\begin{bmatrix}
	\bX_t \\ \bX_{t-1} \\ \vdots \\ \bX_{t-N_t+1} 
\end{bmatrix}	
\right. \,.
$$
Each of these input features are then processes through 2 fully connected layers with $32$ units each before being converted to the law parameters $\Theta$
by the output layer. 
A representation of this simple neural network model is given in Fig. \ref{fig_dense} below.
The  probabilistic prediction results for both Corsica and The Netherlands stations for this model are reported  in tables  \ref{table_D1_A} for a prediction horizon $h=1$ hour and 
table \ref{table_D1_B} for $h=6$ hours.  As one can see, all the obtained scores are slightly worse than the ones reported in Sec. \ref{sec:results} obtained with the LSTM model. However, all the observations and comments remain qualitatively the sames and in particular one can see that the M-Rice distribution provides the better performances overall.
\begin{figure}[h!]
	\centering
	\hspace*{-0.5cm}
	\includegraphics[width=0.6\columnwidth]{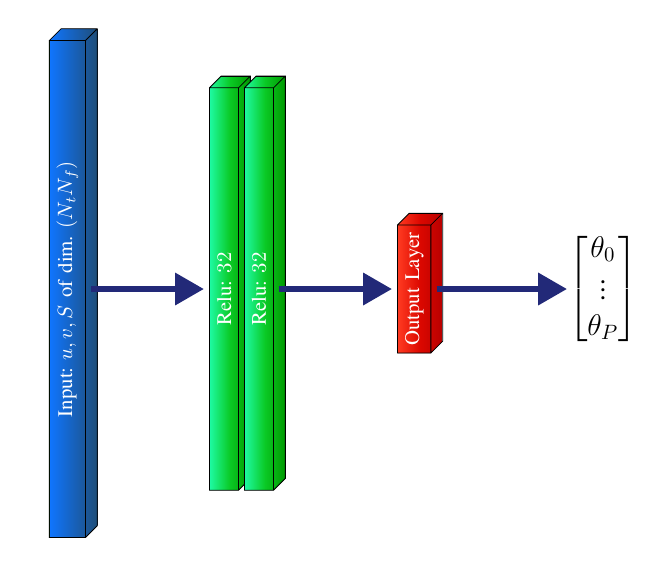}
	\caption{Representation of the fully connected model we tested as an alternative to the LSTM architecture. The 2 green blocks represent fully connected layers with a "Relu" activation function.}
	\label{fig_dense}
\end{figure}

\begin{table}[H]%
		\resizebox{0.5\textwidth}{!}{ %
	\begin{tabular}{c}%
		Corsica\\%
		\\%
		\begin{tabular}{|l|rr|rr|}
			\hline
			Law       &   logScore &   CRPS &    CSL &     twCRPS \\
			\hline
			TNormal   &     1.3007 & 0.5355 & 0.1564 & 0.0465 \\
			Weibull   &     1.3110 & 0.5410 & 0.1610 & 0.0465 \\
			Lognormal &     1.3722 & 0.5453 & 0.1611 & 0.0481\\
			Gamma     &     1.3614 & 0.5506 & 0.1662 & 0.0495 \\
			Nakagami  &     1.3124 & 0.5452 & 0.1609 & 0.0507\\
			Rice      &     1.2988 & 0.5356 & 0.1599 & \textbf{0.0461} \\
			M-Rice  &     \textbf{1.2772} & \textbf{0.5324} & \textbf{0.1555} & 0.0462 \\
			Rayleigh-Rice     &     1.3280 & 0.5441 & 0.1688 & 0.0477\\
			
			\hline
		\end{tabular}\\%
	\end{tabular}}%
	\resizebox{0.5\textwidth}{!}{ %
	\begin{tabular}{c}%
		The Netherlands\\%
		\\%
		\begin{tabular}{|l|rr|rr|}
			\hline
			Law       &   logScore &   CRPS &    CSL &     twCRPS \\
			\hline
			TNormal   &     1.1555 & 0.4402 & 0.1282 & \textbf{0.0288} \\
			Weibull   &     1.1672 & 0.4442 & 0.1333 & 0.0295 \\
			Lognormal &     1.1850 & 0.4422 & 0.1294 & 0.0291 \\
			Gamma     &     1.1774 & 0.4454 & 0.1335 & 0.0296 \\
			Nakagami  &     1.1530 & 0.4418 & 0.1309 & 0.0291 \\
			Rice      &     1.1547 & 0.4407 & 0.1299 &  \textbf{0.0288} \\
			M-Rice  &     \textbf{1.1357} & \textbf{0.4388} & \textbf{0.1275} & 0.0290 \\
			Rayleigh-Rice     &     1.1553 & 0.4412 & 0.1297 & 0.0289\\
			
			\hline
		\end{tabular}\\%
	\end{tabular}}%
	\caption{Score values obtained with the fully connected neural network model. Scores have been averaged over the four considered sites in Corsica (left) and the four considered sites in the Netherlands (right), for $h=1$ hour and $N=15$.  }%
	\label{table_D1_A}
\end{table}

\begin{table}[H]%
		\resizebox{0.5\textwidth}{!}{ %
			\begin{tabular}{c}%
			Corsica\\%
		\\%
		\begin{tabular}{|l|rr|rr|}
			\hline
			Law       &   logScore &   CRPS &    CSL &     twCRPS \\
			\hline
			TNormal   &     1.5982 & 0.7440 & 0.2221 & 0.0700 \\
			Weibull   &     1.5895 & \textbf{0.7438} & 0.2208 & \textbf{0.0696} \\
			Lognormal &     1.6617 & 0.7649 & 0.2301 & 0.0741 \\
			Gamma     &     1.6150 & 0.7500 & \textbf{0.2200} & 0.0715 \\
			Nakagami  &     1.5962 & 0.7478 & 0.2209 & 0.0706 \\
			Rice      &     1.6290 & 0.7586 & 0.2286 & 0.0706 \\
			M-Rice  &     \textbf{1.5803} & 0.7577 & 0.2274 & 0.0726 \\
			Rayleigh-Rice     &     1.6304 & 0.7615 & 0.2261 & 0.0711 \\
			
			\hline
		\end{tabular}\\%
	\end{tabular}}%
	\resizebox{0.5\textwidth}{!}{ %
	\begin{tabular}{c}%
		The Netherlands\\%
		\\%
		\begin{tabular}{|l|rr|rr|}
			\hline
			Law       &   logScore &   CRPS &    CSL &     twCRPS \\
			\hline
			TNormal   &     1.6191 & \textbf{0.7112} & 0.2108 &  \textbf{0.0513} \\
			Weibull   &     1.6144 & 0.7144 & 0.2092 & \textbf{0.0513} \\
			Lognormal &     1.7154 & 0.7538 & 0.2243 & 0.0535 \\
			Gamma     &     1.6311 & 0.7239 & 0.2115 & 0.0524 \\
			Nakagami  &     1.6095 & 0.7146 & 0.2088 & 0.0520 \\
			Rice      &     1.6153 & 0.7156 & 0.2133 & 0.0518 \\
			M-Rice  &     \textbf{1.6016} & 0.7138 & \textbf{0.2063} &  \textbf{0.0513}\\
			Rayleigh-Rice     &     1.6169 & 0.7161 & 0.2117 & 0.0514\\
			
			\hline
		\end{tabular}\\%
	\end{tabular}}%
	\caption{Score values obtained with the fully connected neural network model. Scores have been averaged over the four considered sites in Corsica (left) and the four considered sites in the Netherlands (right), for $h=6$ hour and $N=15$.  }%
	\label{table_D1_B}
\end{table}

\subsection{Regime switching linear model inspired from \cite{gneiting_wind_speed_forecasting_2006}} 

\begin{figure}[h]
	\centering
	\hspace*{-0.5cm}
	\includegraphics[width=0.75\columnwidth]{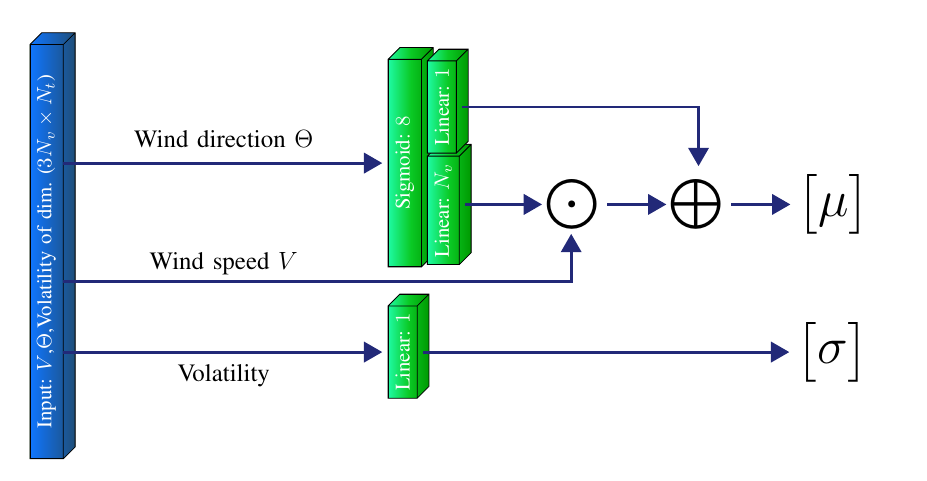}
	\caption{Sketch of the RST network built as an extension of the reference model introduced by Gneiting et al. in \cite{gneiting_wind_speed_forecasting_2006}. Green boxes stand for fully connected layers, $\bigodot$ for a matrix multiplication layer and $\bigoplus$ for a sum layer. The outputs are the parameters $\mu$ and $\sigma$ of a truncated normal distribution.}
	\label{fig_rst}
\end{figure}

The "Regime Switching Spatio Temporal" model introduced in \cite{gneiting_wind_speed_forecasting_2006}  has been designed for the specific situation
considered in the paper with two main wind regimes and two neighbor stations surrounding a reference site.  The RST model simply consists in forecasting the parameters $\mu$ and $\sigma$ of a truncated normal law as respectively a linear regression over past wind speed values for $\mu$  and as the mean value of the square velocity variations (referred to as the "volatility") for $\sigma$. The ``regime switching" relies in the chosen values of the regression coefficients for $\mu$ that depends on the
wind direction observed at the two neighbors sites of the station considered in the study. 
The RST model cannot be directly used in our context and 
in order to adapt to more general situations than the peculiar one considered in \cite{gneiting_wind_speed_forecasting_2006}, i.e., to an arbitrary number of neighboring sites with any number of possible wind direction regimes, we have considered the following variant: The vector of past and present wind directions at the considered stations and also at all the neighboring stations is provided as input of a 2-layer network of fully connected units. The output of these layers consists in the set of of coefficients of the linear regression of input past wind speeds at the reference site and its neighbor in order to predict the location  parameter $\mu$ of the considered conditional wind speed distribution. The variance parameter $\sigma^2$ is build, along the same line as in the original model of Gneiting et al., i.e., as proportional to the mean observed "volatility", the proportionality factor being learnable
during the training process.  The modified RST model is depicted in Fig. \ref{fig_rst}.
As for all neural network models, the RST model parameters are chosen in order to optimize a log-likelihood score function.


 The performances of the RST model average performances for stations in Corsica (left table) and the Netherlands (right tables). LogScore and CRPS computed using RST method are reported at horizons 1h and 6h. table \ref{table_rst} where, along the same line as Table \ref{table1}, results have been averaged for all the stations in Corsica on one hand and all the stations in the Netherlands on the other hand. 
 As for the fully connected neural network model described above and in Section \ref{sec:pdf_comparison}, we considered the wind speeds and directions of the target station and its 15 closest neighbors as input, extending up to $3 h+ 1$ hours into the past  (where $h$ represents the forecast horizon).  When one compares the performances of the RST model and its Neural Network counterparts, one can see that the latter, particularly the LSTM version, provide significantly better results. Notably one can remark that the truncated normal law within the LSTM neural network model provides rather good performances while it leads to significantly worse scores within  the "Regime Switching" linear model similar the one proposed in \cite{gneiting_wind_speed_forecasting_2006}.
 
 \begin{table}[h!]%
    \scriptsize
 	\resizebox{0.5\textwidth}{!} {%
 		\begin{tabular}{c}%
 			Corsica\\%
 			\begin{tabular}{|l|c|c|}
 				\hline
 				\diagbox[width=7em]{horizon}{Score}     &   LogS &   CRPS  \\
 				\hline
 				1h      &      1.34 &  0.74 \\
 				6h &      1.69 &  1.09   \\
 				\hline
 			\end{tabular}\\
 		\end{tabular}%
 	}%
 	\resizebox{0.5\textwidth}{!} {%
 		\begin{tabular}{c}%
 			The Netherlands\\%
 			\begin{tabular}{|l|c|c|c|c|}
 				\hline
 				\diagbox[width=7em]{horizon}{Score}     &   LogS &   CRPS \\
 				\hline
 				1 h      &     1.15  & 0.63 \\
 				6 h &     1.63 & 1.04  \\
 				\hline
 			\end{tabular}\\%
 		\end{tabular}%
 	}%
 	\caption{RST model performance at horizons $h=1$ and $h=6$ hours when one considers wind speed information from  the reference site and  from $N=15$ neighboring stations. The reported LogScore and CRPS are the average values over the validation period for all stations in Corsica (left table) and the Netherlands (right table). When one compares these results to those of Table \ref{table1}, one can see that the RST model is significantly less performing than LSTM model with all the tested probability laws.}%
 	\label{table_rst}
 \end{table}

\newpage

\begin{thebibliography}{10}
	\expandafter\ifx\csname url\endcsname\relax
	\def\url#1{\texttt{#1}}\fi
	\expandafter\ifx\csname urlprefix\endcsname\relax\def\urlprefix{URL }\fi
	\expandafter\ifx\csname href\endcsname\relax
	\def\href#1#2{#2} \def\path#1{#1}\fi
	
	\bibitem{Wang2021}
	Y.~Wang, R.~Zou, F.~Liu, L.~Zhang, Q.~Liu, A review of wind speed and wind
	power forecasting with deep neural networks, Applied Energy 304 (2021)
	117766.
	\newblock \href
	{https://doi.org/https://doi.org/10.1016/j.apenergy.2021.117766}
	{\path{doi:https://doi.org/10.1016/j.apenergy.2021.117766}}.
	
	\bibitem{7764085}
	Y.~Mao, W.~Shaoshuai, A review of wind power forecasting and prediction, in:
	2016 International Conference on Probabilistic Methods Applied to Power
	Systems (PMAPS), 2016, pp. 1--7.
	\newblock \href {https://doi.org/10.1109/PMAPS.2016.7764085}
	{\path{doi:10.1109/PMAPS.2016.7764085}}.
	
	\bibitem{hashli20}
	S.~Hanifi, X.~Liu, Z.~Lin, S.~Lotfian, A critical review of wind power
	forecasting methods-past, present and future, Energies 13 (2020) 3764.
	\newblock \href {https://doi.org/10.3390/en13153764}
	{\path{doi:10.3390/en13153764}}.
	
	\bibitem{5619586}
	S.~S. Soman, H.~Zareipour, O.~Malik, P.~Mandal, A review of wind power and wind
	speed forecasting methods with different time horizons, in: North American
	Power Symposium 2010, 2010, pp. 1--8.
	\newblock \href {https://doi.org/10.1109/NAPS.2010.5619586}
	{\path{doi:10.1109/NAPS.2010.5619586}}.
	
	\bibitem{GIEBEL201759}
	G.~Giebel, G.~Kariniotakis,
	\href{https://www.sciencedirect.com/science/article/pii/B9780081005040000032}{3
		- wind power forecasting - a review of the state of the art}, in:
	G.~Kariniotakis (Ed.), Renewable Energy Forecasting, Woodhead Publishing
	Series in Energy, Woodhead Publishing, 2017, pp. 59--109.
	\newblock \href
	{https://doi.org/https://doi.org/10.1016/B978-0-08-100504-0.00003-2}
	{\path{doi:https://doi.org/10.1016/B978-0-08-100504-0.00003-2}}.
	\newline\urlprefix\url{https://www.sciencedirect.com/science/article/pii/B9780081005040000032}
	
	\bibitem{kapisigi04}
	G.~Kariniotakis, P.~Pinson, N.~Siebert, G.~Giebel, R.~Barthelmie, The state of
	the art in short term prediction of wind power - from an offshore
	perspective, in: Proceedings of 2004 SeaTechWeek, Brest, France, 2004.
	
	\bibitem{Pinson2013}
	P.~Pinson, {Wind Energy: Forecasting Challenges for Its Operational
		Management}, Statistical Science 28~(4) (2013) 564 -- 585.
	\newblock \href {https://doi.org/10.1214/13-STS445}
	{\path{doi:10.1214/13-STS445}}.
	
	\bibitem{ReviewWindProbaForecasting2014}
	Y.~Zhang, J.~Wang, X.~Wang, Review on probabilistic forecasting of wind power
	generation, Renewable and Sustainable Energy Reviews 32 (2014) 255--270.
	\newblock \href {https://doi.org/https://doi.org/10.1016/j.rser.2014.01.033}
	{\path{doi:https://doi.org/10.1016/j.rser.2014.01.033}}.
	
	\bibitem{xie_overview_2023}
	Y.~Xie, C.~Li, M.~Li, F.~Liu, M.~Taukenova,
	\href{https://linkinghub.elsevier.com/retrieve/pii/S2589004222020776}{An
		overview of deterministic and probabilistic forecasting methods of wind
		energy}, iScience 26~(1) (2023) 105804.
	\newblock \href {https://doi.org/10.1016/j.isci.2022.105804}
	{\path{doi:10.1016/j.isci.2022.105804}}.
	\newline\urlprefix\url{https://linkinghub.elsevier.com/retrieve/pii/S2589004222020776}
	
	\bibitem{gneiting_Ensemble_05}
	T.~Gneiting, A.~E. Raftery,
	\href{https://www.science.org/doi/abs/10.1126/science.1115255}{Weather
		forecasting with ensemble methods}, Science 310~(5746) (2005) 248--249.
	\newblock \href
	{http://arxiv.org/abs/https://www.science.org/doi/pdf/10.1126/science.1115255}
	{\path{arXiv:https://www.science.org/doi/pdf/10.1126/science.1115255}}, \href
	{https://doi.org/10.1126/science.1115255}
	{\path{doi:10.1126/science.1115255}}.
	\newline\urlprefix\url{https://www.science.org/doi/abs/10.1126/science.1115255}
	
	\bibitem{Quantile_Regression_05}
	R.~Koenker, K.~F. Hallock,
	\href{https://www.aeaweb.org/articles?id=10}{Quantile regression}, Journal of
	Economic Perspectives 15~(4) (2001) 143--156.
	\newblock \href {https://doi.org/10.1257/jep.15.4.143}
	{\path{doi:10.1257/jep.15.4.143}}.
	\newline\urlprefix\url{https://www.aeaweb.org/articles?id=10}
	
	\bibitem{Nielsen_06}
	H.~A. Nielsen, H.~Madsen, T.~S. Nielsen,
	\href{https://onlinelibrary.wiley.com/doi/abs/10.1002/we.180}{Using quantile
		regression to extend an existing wind power forecasting system with
		probabilistic forecasts}, Wind Energy 9~(1-2) (2006) 95--108.
	\newblock \href
	{http://arxiv.org/abs/https://onlinelibrary.wiley.com/doi/pdf/10.1002/we.180}
	{\path{arXiv:https://onlinelibrary.wiley.com/doi/pdf/10.1002/we.180}}, \href
	{https://doi.org/https://doi.org/10.1002/we.180}
	{\path{doi:https://doi.org/10.1002/we.180}}.
	\newline\urlprefix\url{https://onlinelibrary.wiley.com/doi/abs/10.1002/we.180}
	
	\bibitem{HE2018374}
	Y.~He, H.~Li,
	\href{https://www.sciencedirect.com/science/article/pii/S0196890418302255}{Probability
		density forecasting of wind power using quantile regression neural network
		and kernel density estimation}, Energy Conversion and Management 164 (2018)
	374--384.
	\newblock \href
	{https://doi.org/https://doi.org/10.1016/j.enconman.2018.03.010}
	{\path{doi:https://doi.org/10.1016/j.enconman.2018.03.010}}.
	\newline\urlprefix\url{https://www.sciencedirect.com/science/article/pii/S0196890418302255}
	
	\bibitem{bamu_23}
	R.~Baïle, J.-F. Muzy,
	\href{https://www.sciencedirect.com/science/article/pii/S0360544222025300}{Leveraging
		data from nearby stations to improve short-term wind speed forecasts}, Energy
	263 (2023) 125644.
	\newblock \href {https://doi.org/https://doi.org/10.1016/j.energy.2022.125644}
	{\path{doi:https://doi.org/10.1016/j.energy.2022.125644}}.
	\newline\urlprefix\url{https://www.sciencedirect.com/science/article/pii/S0360544222025300}
	
	\bibitem{gneiting_wind_speed_forecasting_2006}
	T.~Gneiting, K.~Larson, K.~Westrick, M.~G. Genton, E.~Aldrich, Calibrated
	probabilistic forecasting at the stateline wind energy center: The
	regime-switching space–time method, Journal of the American Statistical
	Association 101~(475) (2006) 968--979.
	\newblock \href {https://doi.org/10.1198/016214506000000456}
	{\path{doi:10.1198/016214506000000456}}.
	
	\bibitem{Pinson_2012}
	P.~Pinson, Very-short-term probabilistic forecasting of wind power with
	generalized logit–normal distributions, Journal of the Royal Statistical
	Society: Series C (Applied Statistics) 61~(4) (2012) 555--576.
	\newblock \href
	{https://doi.org/https://doi.org/10.1111/j.1467-9876.2011.01026.x}
	{\path{doi:https://doi.org/10.1111/j.1467-9876.2011.01026.x}}.
	
	\bibitem{TastuPinson14}
	J.~Tastu, P.~Pinson, P.-J. Trombe, H.~Madsen, Probabilistic forecasts of wind
	power generation accounting for geographically dispersed information, IEEE
	Transactions on Smart Grid 5~(1) (2014) 480--489.
	\newblock \href {https://doi.org/10.1109/TSG.2013.2277585}
	{\path{doi:10.1109/TSG.2013.2277585}}.
	
	\bibitem{holland_forecasting_2014}
	M.~J. Holland, K.~Ikeda,
	\href{http://ieeexplore.ieee.org/document/6853986/}{Forecasting in wind
		energy applications with site-adaptive {Weibull} estimation}, in: 2014 {IEEE}
	{International} {Conference} on {Acoustics}, {Speech} and {Signal}
	{Processing} ({ICASSP}), IEEE, Florence, Italy, 2014, pp. 2184--2188.
	\newblock \href {https://doi.org/10.1109/ICASSP.2014.6853986}
	{\path{doi:10.1109/ICASSP.2014.6853986}}.
	\newline\urlprefix\url{http://ieeexplore.ieee.org/document/6853986/}
	
	\bibitem{XIANG2020113098}
	L.~Xiang, J.~Li, A.~Hu, Y.~Zhang, Deterministic and probabilistic multi-step
	forecasting for short-term wind speed based on secondary decomposition and a
	deep learning method, Energy Conversion and Management 220 (2020) 113098.
	\newblock \href
	{https://doi.org/https://doi.org/10.1016/j.enconman.2020.113098}
	{\path{doi:https://doi.org/10.1016/j.enconman.2020.113098}}.
	
	\bibitem{ZHU2019111772}
	S.~Zhu, X.~Yuan, Z.~Xu, X.~Luo, H.~Zhang,
	\href{https://www.sciencedirect.com/science/article/pii/S019689041930754X}{Gaussian
		mixture model coupled recurrent neural networks for wind speed interval
		forecast}, Energy Conversion and Management 198 (2019) 111772.
	\newblock \href
	{https://doi.org/https://doi.org/10.1016/j.enconman.2019.06.083}
	{\path{doi:https://doi.org/10.1016/j.enconman.2019.06.083}}.
	\newline\urlprefix\url{https://www.sciencedirect.com/science/article/pii/S019689041930754X}
	
	\bibitem{Hossain21}
	M.~A. Hossain, R.~K. Chakrabortty, S.~Elsawah, E.~M. Gray, M.~J. Ryan,
	Predicting wind power generation using hybrid deep learning with
	optimization, IEEE Transactions on Applied Superconductivity 31~(8) (2021)
	1--5.
	\newblock \href {https://doi.org/10.1109/TASC.2021.3091116}
	{\path{doi:10.1109/TASC.2021.3091116}}.
	
	\bibitem{Kou2013}
	P.~Kou, F.~Gao, X.~Guan, Sparse online warped gaussian process for wind power
	probabilistic forecasting, Applied Energy 108 (2013) 410--428.
	\newblock \href
	{https://doi.org/https://doi.org/10.1016/j.apenergy.2013.03.038}
	{\path{doi:https://doi.org/10.1016/j.apenergy.2013.03.038}}.
	
	\bibitem{HU2017}
	J.~Hu, J.~Wang, L.~Xiao, A hybrid approach based on the gaussian process with
	t-observation model for short-term wind speed forecasts, Renewable Energy 114
	(2017) 670--685.
	\newblock \href {https://doi.org/https://doi.org/10.1016/j.renene.2017.05.093}
	{\path{doi:https://doi.org/10.1016/j.renene.2017.05.093}}.
	
	\bibitem{HENG2022118029}
	J.~Heng, Y.~Hong, J.~Hu, S.~Wang, Probabilistic and deterministic wind speed
	forecasting based on non-parametric approaches and wind characteristics
	information, Applied Energy 306 (2022) 118029.
	\newblock \href
	{https://doi.org/https://doi.org/10.1016/j.apenergy.2021.118029}
	{\path{doi:https://doi.org/10.1016/j.apenergy.2021.118029}}.
	
	\bibitem{DeepAR}
	D.~Salinas, V.~Flunkert, J.~Gasthaus, T.~Januschowski, Deepar: Probabilistic
	forecasting with autoregressive recurrent networks, International Journal of
	Forecasting 36~(3) (2020) 1181--1191.
	\newblock \href
	{https://doi.org/https://doi.org/10.1016/j.ijforecast.2019.07.001}
	{\path{doi:https://doi.org/10.1016/j.ijforecast.2019.07.001}}.
	
	\bibitem{MASHLAKOV2021116405}
	A.~Mashlakov, T.~Kuronen, L.~Lensu, A.~Kaarna, S.~Honkapuro,
	\href{https://www.sciencedirect.com/science/article/pii/S0306261920317748}{Assessing
		the performance of deep learning models for multivariate probabilistic energy
		forecasting}, Applied Energy 285 (2021) 116405.
	\newblock \href
	{https://doi.org/https://doi.org/10.1016/j.apenergy.2020.116405}
	{\path{doi:https://doi.org/10.1016/j.apenergy.2020.116405}}.
	\newline\urlprefix\url{https://www.sciencedirect.com/science/article/pii/S0306261920317748}
	
	\bibitem{Afrasiabi21}
	M.~Afrasiabi, M.~Mohammadi, M.~Rastegar, S.~Afrasiabi, Advanced deep learning
	approach for probabilistic wind speed forecasting, IEEE Transactions on
	Industrial Informatics 17~(1) (2021) 720--727.
	\newblock \href {https://doi.org/10.1109/TII.2020.3004436}
	{\path{doi:10.1109/TII.2020.3004436}}.
	
	\bibitem{BaMuPo10c}
	R.~{Ba\"ile}, J.~F. Muzy, P.~Poggi, An {M}-{R}ice wind speed frequency
	distribution, Wind Energy 14 (2011) 735--748.
	
	\bibitem{BaMuPo11}
	R.~Ba\"{\i}le, J.~F. Muzy, P.~Poggi, Short-term forecasting of surface layer
	wind speed using a continuous random cascade model, Wind Energy 14~(6) (2011)
	719--734.
	\newblock \href {https://doi.org/https://doi.org/10.1002/we.452}
	{\path{doi:https://doi.org/10.1002/we.452}}.
	
	\bibitem{BaMuPo10}
	R.~{Ba\"ile}, J.~F. Muzy, P.~Poggi, Short term forecasting of surface layer
	wind speed using a continuous random cascade model, Wind Energy 14 (2011)
	719--734.
	
	\bibitem{BaMu10}
	R.~{Ba\"ile}, J.~F. Muzy, Spatial intermittency of surface layer wind
	fluctuations at mesoscale range, Physical Review Letters 105 (2010) 254501.
	
	\bibitem{drobinski_2015}
	P.~Drobinski, C.~Coulais, B.~Jourdier, Surface wind-speed statistics modelling:
	Alternatives to the weibull distribution and performance evaluation,
	Boundary-Layer Meteorology 157~(1) (2015) 97--123.
	\newblock \href {https://doi.org/10.1007/s10546-015-0035-7}
	{\path{doi:10.1007/s10546-015-0035-7}}.
	
	\bibitem{ERA5}
	H.~Hersbach, B.~Bell, P.~Berrisford, G.~Biavati, A.~Horányi,
	J.~Muñoz~Sabater, J.~Nicolas, C.~Peubey, R.~Radu, I.~Rozum, D.~Schepers,
	A.~Simmons, C.~Soci, D.~Dee, J.-N. Thépaut, {ERA5} hourly data on single
	levels from 1940 to present, Copernicus {C}limate {C}hange {S}ervice (C3S)
	{C}limate {D}ata {S}tore (CDS) doi:10.24381/cds.adbb2d477 (2023).
	
	\bibitem{Grassi2015}
	S.~Grassi, F.~Veronesi, R.~Martin, Satellite remote sensed data to improve the
	accuracy of statistical models for wind resource assessment, in: European
	Wind EnergyAssociation Annual Conference and Exhibition, 2015.
	
	\bibitem{Amato22}
	
	
	\bibitem{Lopez2022}
	C.~Lopez-Villalobos, O.~Martínez-Alvarado, O.~Rodriguez-Hernandez,
	R.~Romero-Centeno, Analysis of the influence of the wind speed profile on
	wind power production, Energy Reports 8 (2022) 8079--8092.
	\newblock \href {https://doi.org/https://doi.org/10.1016/j.egyr.2022.06.046}
	{\path{doi:https://doi.org/10.1016/j.egyr.2022.06.046}}.
	
	\bibitem{IEA23}
	C.~Möhrlen, J.~W. Zack, G.~Giebel (Eds.), IEA Wind Recommended Practice for
	the Implementation of Renewable Energy Forecasting Solutions, Wind Energy
	Engineering, Academic Press, 2023.
	\newblock \href
	{https://doi.org/https://doi.org/10.1016/B978-0-44-318681-3.00015-5}
	{\path{doi:https://doi.org/10.1016/B978-0-44-318681-3.00015-5}}.
	
	\bibitem{Drechsel12}
	S.~Drechsel, G.~J. Mayr, J.~W. Messner, R.~Stauffer, Wind speeds at heights
	crucial for wind energy: Measurements and verification of forecasts, Journal
	of Applied Meteorology and Climatology 51~(9) (2012) 1602 -- 1617.
	\newblock \href {https://doi.org/https://doi.org/10.1175/JAMC-D-11-0247.1}
	{\path{doi:https://doi.org/10.1175/JAMC-D-11-0247.1}}.
	
	\bibitem{Good16}
	I.~J. Goodfellow, Y.~Bengio, A.~Courville, Deep Learning, MIT Press, Cambridge,
	MA, USA, 2016.
	
	\bibitem{Wu_2016}
	W.~Wu, K.~Chen, Y.~Qiao, Z.~Lu, Probabilistic short-term wind power forecasting
	based on deep neural networks, in: 2016 International Conference on
	Probabilistic Methods Applied to Power Systems (PMAPS), 2016, pp. 1--8.
	\newblock \href {https://doi.org/10.1109/PMAPS.2016.7764155}
	{\path{doi:10.1109/PMAPS.2016.7764155}}.
	
	\bibitem{vaswani2017attention}
	A.~Vaswani, N.~Shazeer, N.~Parmar, J.~Uszkoreit, L.~Jones, A.~N. Gomez,
	{\L}.~Kaiser, I.~Polosukhin, Attention is all you need, Advances in neural
	information processing systems 30 (2017).
	
	\bibitem{review_wind_speed_distributions_09}
	J.~Carta, P.~Ramírez, S.~Velázquez, {A review of wind speed probability
		distributions used in wind energy analysis: Case studies in the Canary
		Islands}, Renewable and Sustainable Energy Reviews 13~(5) (2009) 933--955.
	
	\bibitem{review_wind_Speed_distributions_21}
	H.~Shi, Z.~Dong, N.~Xiao, Q.~Huang, Frontiers in Energy Research 9 (2021).
	\newblock \href {https://doi.org/10.3389/fenrg.2021.769920}
	{\path{doi:10.3389/fenrg.2021.769920}}.
	
	\bibitem{conradsen}
	K.~Conradsen, L.~B. Nielsen, Review of weibull statistics for estimation of
	wind speed distributions, Journal of Applied Meteorology 23 (1984)
	1173--1183.
	
	\bibitem{Dookie18}
	I.~Dookie, S.~Rocke, A.~Singh, C.~Ramlal, Evaluating wind speed probability
	distribution models with a novel goodness of fit metric: a trinidad and
	tobago case study, International Journal of Energy and Environmental
	Engineering 9 (05 2018).
	\newblock \href {https://doi.org/10.1007/s40095-018-0271-y}
	{\path{doi:10.1007/s40095-018-0271-y}}.
	
	\bibitem{yu19}
	J.~Yu, Y.~Fu, Y.~Yu, S.~Wu, Y.~Wu, M.~You, S.~Guo, M.~Li, Assessment of
	offshore wind characteristics and wind energy potential in bohai bay, china,
	Energies 12~(15) (2019).
	\newblock \href {https://doi.org/10.3390/en12152879}
	{\path{doi:10.3390/en12152879}}.
	
	\bibitem{Gugliani2020}
	G.~K. Gugliani, \href{https://doi.org/10.1063/5.0024052}{{Comparison of
			different multi-parameters probability density models for wind resources
			assessment}}, Journal of Renewable and Sustainable Energy 12~(6) (2020)
	063303.
	\newblock \href
	{http://arxiv.org/abs/https://pubs.aip.org/aip/jrse/article-pdf/doi/10.1063/5.0024052/15777103/063303\_1\_online.pdf}
	{\path{arXiv:https://pubs.aip.org/aip/jrse/article-pdf/doi/10.1063/5.0024052/15777103/063303\_1\_online.pdf}},
	\href {https://doi.org/10.1063/5.0024052} {\path{doi:10.1063/5.0024052}}.
	\newline\urlprefix\url{https://doi.org/10.1063/5.0024052}
	
	\bibitem{suriadi21}
	S.~Suriadi, M.~Nabilah, M.~Zainal, M.~Yanis, M.~Marwan, H.~Hafidh, M.~Affan,
	Comparative analysis of wind energy potential with nakagami and weibull
	distribution methods for wind turbine planning 12 (2023) 104–115.
	\newblock \href {https://doi.org/https://doi.org/10.13170/aijst.12.1.30736}
	{\path{doi:https://doi.org/10.13170/aijst.12.1.30736}}.
	
	\bibitem{Rice}
	S.~O. Rice, Mathematical analysis of random noise, Bell System Technical
	Journal 24 (1945) 46--156.
	
	\bibitem{Fri95}
	U.~Frisch, Turbulence, Cambridge Univ. Press, Cambridge, 1995.
	
	\bibitem{abst64}
	M.~{Abramowitz}, I.~A. {Stegun}, Handbook of Mathematical Functions with
	Formulas, Graphs, and Mathematical Tables, ninth dover printing, tenth gpo
	printing Edition, Dover, New York City, 1964.
	
	\bibitem{gneiting2014probabilistic}
	T.~Gneiting, M.~Katzfuss, Probabilistic forecasting, Annual Review of
	Statistics and Its Application 1 (2014) 125--151.
	
	\bibitem{du2021beyond}
	H.~Du, Beyond strictly proper scoring rules: The importance of being local,
	Weather and forecasting 36~(2) (2021) 457--468.
	
	\bibitem{gneiting2007strictly}
	T.~Gneiting, A.~E. Raftery, Strictly proper scoring rules, prediction, and
	estimation, Journal of the American statistical Association 102~(477) (2007)
	359--378.
	
	\bibitem{jolliffe2008impenetrable}
	I.~T. Jolliffe, The impenetrable hedge: A note on propriety, equitability and
	consistency, Meteorological Applications: A journal of forecasting, practical
	applications, training techniques and modelling 15~(1) (2008) 25--29.
	
	\bibitem{good1952rational}
	I.~J. Good, Rational decisions, Journal of the Royal Statistical Society:
	Series B (Methodological) 14~(1) (1952) 107--114.
	
	\bibitem{roulston2002evaluating}
	M.~S. Roulston, L.~A. Smith, Evaluating probabilistic forecasts using
	information theory, Monthly Weather Review 130~(6) (2002) 1653--1660.
	
	\bibitem{diks2011likelihood}
	C.~Diks, V.~Panchenko, D.~Van~Dijk, Likelihood-based scoring rules for
	comparing density forecasts in tails, Journal of Econometrics 163~(2) (2011)
	215--230.
	
	\bibitem{selten1998axiomatic}
	R.~Selten, Axiomatic characterization of the quadratic scoring rule,
	Experimental Economics 1 (1998) 43--61.
	
	\bibitem{stael1970family}
	C.-A.~S. Sta{\"e}l~von Holstein, A family of strictly proper scoring rules
	which are sensitive to distance, Journal of Applied Meteorology and
	Climatology 9~(3) (1970) 360--364.
	
	\bibitem{brown1974admissible}
	T.~A. Brown, Admissible scoring systems for continuous distributions. (1974).
	
	\bibitem{hersbach2000decomposition}
	H.~Hersbach, Decomposition of the continuous ranked probability score for
	ensemble prediction systems, Weather and Forecasting 15~(5) (2000) 559--570.
	
	\bibitem{todter2012generalization}
	J.~T{\"o}dter, B.~Ahrens, Generalization of the ignorance score: Continuous
	ranked version and its decomposition, Monthly Weather Review 140~(6) (2012)
	2005--2017.
	
	\bibitem{gneiting2007probabilistic}
	T.~Gneiting, F.~Balabdaoui, A.~E. Raftery, Probabilistic forecasts, calibration
	and sharpness, Journal of the Royal Statistical Society Series B: Statistical
	Methodology 69~(2) (2007) 243--268.
	
	\bibitem{brocker2009reliability}
	J.~Br{\"o}cker, Reliability, sufficiency, and the decomposition of proper
	scores, Quarterly Journal of the Royal Meteorological Society: A journal of
	the atmospheric sciences, applied meteorology and physical oceanography
	135~(643) (2009) 1512--1519.
	
	\bibitem{diebold1997evaluating}
	F.~X. Diebold, T.~A. Gunther, A.~Tay, Evaluating density forecasts (1997).
	
	\bibitem{delle2006probabilistic}
	L.~Delle~Monache, J.~P. Hacker, Y.~Zhou, X.~Deng, R.~B. Stull, Probabilistic
	aspects of meteorological and ozone regional ensemble forecasts, Journal of
	Geophysical Research: Atmospheres 111~(D24) (2006).
	
	\bibitem{pinson2007non}
	P.~Pinson, H.~A. Nielsen, J.~K. M{\o}ller, H.~Madsen, G.~N. Kariniotakis,
	Non-parametric probabilistic forecasts of wind power: required properties and
	evaluation, Wind Energy: An International Journal for Progress and
	Applications in Wind Power Conversion Technology 10~(6) (2007) 497--516.
	
	\bibitem{lerch2017forecaster}
	S.~Lerch, T.~L. Thorarinsdottir, F.~Ravazzolo, T.~Gneiting, Forecaster's
	dilemma: extreme events and forecast evaluation, Statistical Science (2017)
	106--127.
	
	\bibitem{ranjan2010combining}
	R.~Ranjan, T.~Gneiting, Combining probability forecasts, Journal of the Royal
	Statistical Society Series B: Statistical Methodology 72~(1) (2010) 71--91.
	
	\bibitem{marqum_approx}
	M.~Bocus, C.~Dettmann, J.~Coon, An approximation of the first order marcum
	$q$-function with application to network connectivity analysis, IEEE
	Communications Letters 17 (10 2012).
	\newblock \href {https://doi.org/10.1109/LCOMM.2013.011513.122462}
	{\path{doi:10.1109/LCOMM.2013.011513.122462}}.
	
	\bibitem{Pedregosa:2011}
	F.~Pedregosa, Scikit-learn: Machine learning in python, J. Mach. Learn. Res. 12
	(2011) 2825--2830.
	
	\bibitem{candille2005evaluation}
	G.~Candille, O.~Talagrand, Evaluation of probabilistic prediction systems for a
	scalar variable, Quarterly Journal of the Royal Meteorological Society: A
	journal of the atmospheric sciences, applied meteorology and physical
	oceanography 131~(609) (2005) 2131--2150.
	
\end{thebibliography}

\end{document}